\documentclass{article}
\usepackage[utf8]{inputenc}
\usepackage{amsmath,amsthm,amsfonts,amssymb}
\usepackage{xcolor}
\usepackage{graphicx,tikz,setspace}
\usepackage[margin=2.5cm]{geometry}
\usepackage[colorlinks=true]{hyperref}
\usepackage{float}
\usepackage{mathtools}
\usepackage{subcaption}
\usepackage{caption}
\usepackage{enumitem}
\usepackage{indentfirst}
\usetikzlibrary{positioning}
\usetikzlibrary{topaths,calc}
\usepackage{here}
\usepackage{aliascnt}
\usepackage{caption}
\usepackage[backend=biber,style=alphabetic, maxnames=10]{biblatex}
\usepackage{csquotes}
\usepackage{harpoon}

\usepackage[ruled, lined, linesnumbered, commentsnumbered, longend]{algorithm2e}

\usepackage[english]{babel}  
\addto\extrasenglish{}
\addto\extrasenglish{}
\addto\extrasenglish{}

\usepackage[rm]{titlesec}
\titleformat*{\section}{\center\large\bfseries}
\titleformat*{\subsection}{\center\bfseries}
\titleformat*{\subsubsection}{\center\bfseries}

\hypersetup{
colorlinks=true,
citecolor=blue,
linkcolor=red,
urlcolor=cyan}

\usepackage{fancyhdr}
\usepackage{lastpage}

\usepackage{tikz}
\usetikzlibrary{arrows, shapes, fit, arrows.meta, bending, shapes.geometric, shapes.callouts}

\tikzstyle{io} = [trapezium, trapezium stretches=true, trapezium left angle=70, trapezium right angle=110, minimum width=3cm, minimum height=2em, text centered, draw=black]
\tikzstyle{process} = [rectangle, draw=black, text centered, minimum width=3cm, minimum height=2em, minimum height=2em]
\tikzstyle{decision} = [diamond, aspect=2, draw=black, text centered,minimum width=3cm, minimum height=2em, minimum height=2em]
\tikzstyle{arrow} = [thick,->,>=stealth]

\usepackage{pgfplots}
\usepackage{color}
\pgfplotsset{compat=newest}
\usetikzlibrary{arrows.meta}
\usepgfplotslibrary{fillbetween}

\pgfarrowsdeclare{mylatex'}{mylatex'}
{
  \newdimen\len
  \len=\pgfgetarrowoptions{mylatex'}
  \pgfarrowsleftextend{-0.4\len}
  \pgfarrowsrightextend{0.6\len}
}
{
  \newdimen\len
  \len=\pgfgetarrowoptions{mylatex'}
  \pgfpathmoveto{\pgfqpoint{0.6\len}{0\len}}
  \pgfpathcurveto
  {\pgfqpoint{0.35\len}{0.05\len}}
  {\pgfqpoint{-0.1\len}{0.15\len}}
  {\pgfqpoint{-0.4\len}{0.375\len}}
  \pgfpathcurveto
  {\pgfqpoint{-0.15\len}{0.1\len}}
  {\pgfqpoint{-0.15\len}{-0.1\len}}
  {\pgfqpoint{-0.4\len}{-0.375\len}}
  \pgfpathcurveto
  {\pgfqpoint{-0.1\len}{-0.15\len}}
  {\pgfqpoint{0.35\len}{-0.05\len}}
  {\pgfqpoint{0.6\len}{0\len}}
  \pgfusepathqfill
}
\pgfsetarrowoptions{mylatex'}{8pt}
\pgfkeys{/tiplen/.default=8pt, /tiplen/.code={\pgfsetarrowoptions{mylatex'}{#1}}}
\tikzset{
    myarrow/.style={-{Triangle[length=3mm,width=1mm]}}
}

\numberwithin{equation}{section}
\theoremstyle{definition}
\newtheorem{definition}{Definition}[section]

\newaliascnt{example}{definition} 
\newtheorem{example}[example]{Example}
\aliascntresetthe{example}

\newaliascnt{remark}{definition} 
\newtheorem{remark}[remark]{Remark}
\aliascntresetthe{remark}

\newaliascnt{prob}{definition} 
\newtheorem{prob}[prob]{Problem}
\aliascntresetthe{prob}

\newcommand{\N}{\mathbb{N}}

\newcommand{\R}{\mathbb{R}}

\def\:={\coloneqq} 

\def\kakko<#1>{\left\langle #1 \right\rangle}
\def\diam(#1){\mathsf{diam}(#1)}
\def\W(#1){W_1\big(#1\big)}
\def\wh(#1){W_h\big(#1\big)}
\def\conv(#1){\textrm{conv}\left( #1 \right)}
\def\01{\{0,1\}}
\def\L(#1){#1\textrm{-Lip}}
\def\w-#1-Lip{\textrm{w-}$#1$\textrm{-Lip}}
\def\3|{|\hspace{-0.4mm}|\hspace{-0.4mm}|}
\def\Lip(#1){\textsf{Lip}_w^{#1}(V)}

\usepackage{xcolor}

\newcommand{\seq}{\subseteq} 

\newcommand{\fa}{\ \forall \ }

\DeclarePairedDelimiter\abs{\lvert}{\rvert}%
\DeclarePairedDelimiter\no{\lVert}{\rVert}%

\newcommand{\ar}[1]{{\overrightharp{#1}}} 

\newcommand{\variable}[1]{\textbf{#1}}

\makeatletter
\def\blfootnote{\xdef\@thefnmark{}\@footnotetext}
\makeatother

\title{Mitsubishi-A Final Paper}
\author{}
\date{}

\begin{document}

\begin{titlepage}

\newcommand{\HRule}{\rule{\linewidth}{0.5mm}} 

\center 
 



\HRule \\[0.4cm]
{ \huge \bfseries Two Online Map Matching Algorithms Based on Analytic Hierarchy Process and Fuzzy Logic}\\[0.4cm] 
\HRule \\[1.5cm]
 

\newcommand{\mycomment}[1]{}

\begin{minipage}{0.4\textwidth}
\begin{flushleft} \large
\emph{Authors:}\\
  \textsc{Jeremy J. Lin}\textsuperscript{1}\\
  \textsc{Tomoro Mochida}\textsuperscript{2}\\
  \textsc{Riley C. W. O'Neill}\textsuperscript{3}\\
  \textsc{Atsuro Yoshida}\textsuperscript{4}\\
  \textsc{Masashi Yamazaki Kanazawa}\textsuperscript{5} \\
  \textsc{Akinobu Sasada}\textsuperscript{5} \\
\end{flushleft}
\end{minipage}
\vfill
\center\begin{minipage}{0.35\textwidth}
\begin{flushleft}\small
Institute\\
\textsuperscript{1} University of California Irvine\\
\textsuperscript{2} Tohoku University\\
\textsuperscript{3} University of Minnesota, Twin Cities\\
\textsuperscript{4} Nagoya University\\
\textsuperscript{5} Advanced Technology R\&D Center, Mitsubishi Electric Corporation \\

\end{flushleft}
\end{minipage}\\[2cm]



{\large \today}\\[2cm] 


 
\blfootnote{A. Y. was a recipient of a scholarship from Iwadare Scholarship Foundation in the academic year 2023.}
\vfill 
\end{titlepage}

\tableofcontents \newpage

\section{Introduction}

Map matching is the process of determining the correct traveling route taken by a person or a vehicle using a road network (map) and trajectory data obtained from GPS or other positioning sensors. While it may be easy for humans to estimate, it is difficult to implement as an algorithm. Handling GPS errors and route selection at junctions and parallel roads well is the key to accurate map matching.
Various map matching algorithms have been provided so far and they are typically classified into four types depending on the method used: geometric, topological, probabilistic, and advanced map matching. Each of these methods has its advantages and disadvantages in terms of accuracy, implementation, generalizability, assumptions, etc.
We will explain these methods in detail in Section~\ref{sec:background}.

Our aim of this paper is to develop new map matching algorithms and to improve upon previous work. We address two key approaches: Analytic Hierarchy Process (AHP) map matching and fuzzy logic map matching.
AHP is a decision-making method that combines mathematical analysis with human judgment, and fuzzy logic is an approach to computing based on the degree of truth and aims at modeling the imprecise modes of reasoning from 0 to 1 rather than the usual boolean logic.
Of these algorithms, the way of our applying AHP to map matching is newly developed in this paper, meanwhile, our application of fuzzy logic to map matching is mostly the same as existing research except for some small changes.
Because of the common characteristic that both methods are designed to handle imprecise information and simplicity for implementation, we decided to use these methods.
The details of these methods are discussed in Subsection~\ref{Fuzzy} and Section~\ref{sec:map-matching-algorithms}.

We implemented and tested these methods, comparing their performance with an existing algorithm.
In contrast to previous approaches, our algorithms are designed for ease of implementation, while still achieving adequate performance in terms of both error rate and speed.
On the other hand, several points for improvement were left as future tasks: using better estimation for additional information that the AHP and fuzzy logic map matching algorithms utilize, optimizing parameters that are needed for our map matching algorithm, implementing our algorithms in a faster programming language, and incorporating post-processing after a map matching algorithm.

\subsection*{Acknowledgements}
This research was conducted during g-RIPS (Graduate-level Research in Industrial Projects for Students)-Sendai 2023 program hosted by the Advanced Institute for Materials Research at Tohoku University in collaboration with the Institute for Pure \& Applied Mathematics of the University of California, Los Angeles.
In the program, we focused on the issue that we address in this paper, which was designed in collaboration with our industrial partner, Mitsubishi Electric.
We are deeply grateful for the excellent environment and support provided there, and the financial support from Advanced Technology R\&D Center, Mitsubishi Electric Corporation.
Finally, we would like to express our gratitude to Shunsuke Kano for his helpful comments.

\section{Notation, Problem Summary, and Statement}

\subsection{Notation}
We formulate a graph-theoretic formulation of roads as follows:
\begin{definition}
Graph - a set of points $V = (v_1,v_2,\dots,v_n) \seq \R^d$ equipped with a list of edges $E = \{(e_{i,1},e_{i,2})\}_{i=1}^m \seq \N_n \times \N_n$ ($\N_n = (1,2,...n)$) which denotes connected pairs of points, i.e. $v_{e_{i,1}}$ connects to $v_{e_{i,2}}$ for all $1\leq i \leq m$. We assume $e_{i,1} \neq e_{i,2} \fa i$ to avoid storing needless information. 
\end{definition}

\begin{definition} Vertex / Node - a point in a graph. We use these terms interchangeably. 
\end{definition}

\begin{definition}
Road Network - a directed graph $G$ representing roads under the geometric realization. We leave this deliberately broad, possibly including both junction points and intermediary points:
\end{definition}

\begin{definition} Junction point - a vertex $p_i$ in a road network with a degree of connectivity $d_i$ that satisfies $d_i = 1$ or $d_i \geq 3 \fa i=1,N$. Intuitively, this is a point where one's trajectory can or must change (without taking an illegal u-turn); while the degree 1 points are not "junctions," these are still important points that must be considered when examining the connectivity of the map graph. 
\end{definition}

\begin{definition} Intermediary point - a vertex in a graphical representation of a road with a degree connectivity 2. Intuitively, this is a point where one's trajectory cannot change (without taking an illegal u-turn or exiting the road network. 
\end{definition}

Define the road segment and map adjacency graph as follows:

\begin{definition}
Road segment - a directed subgraph of the road network where the two end nodes $n_1$ and $n_N$ are junction points (i.e. a degree of connectivity that satisfies $d_i = 1$ or $d_i \geq 3 \fa i=1,N$), and all intermediary nodes between them are strictly degree 2 ($d_i = 2 \fa 1<i<N$).
\end{definition}
This is simply to say that if one is on a road segment, one cannot alter one's trajectory unless one travels to an end of the segment (unless taking a U-turn). We leave U-turns to another matter of pre/postprocessing. 

We also define 
\begin{definition}
Map adjacency graph - an abstraction of the road network that encodes the connectivity of road segments - i.e., what path decisions one has when driving or walking. Each edge is a road segment. 
\end{definition}

\subsection{Problem Summary and Statement}

Let us fix $d \geq 2$ (but almost everywhere we consider the case $d \in\{2, 3\}$).

\begin{definition}[Trajectory] \label{Tr}
A \textbf{trajectory} $\mathrm{T}$ is a timeseries of points $P = (p_1,p_2,\dots, p_n) \seq \R^d$ equipped with several features $F = (f_1,f_2,\dots, f_n) \seq \R^k$, where $k\geq 1$. We mandate that $F$ at least includes timestamps, but could possibly include additional attributes, as explained thus:
\begin{enumerate}
    \item Timestamp - a sequence $\mathrm{T}_t = (t_1,t_2,\dots,t_n) \seq \R_+ = [0,\infty)$ that is strictly increasing ($t_1<t_2<\dots <t_n$).
    \item Speed (optional) - a sequence $\mathrm{T}_s = (v_1, v_2,\dots, v_n) \seq \R_+$.
    \item Direction of travel/head (optional) - a sequence of unit vectors $\mathrm{T}_u = (u_1, \dots, u_n) \seq \mathbb{S}^d$ which are the unit velocity vectors ("direction") of each point in the trajectory. 
    \item Acceleration (optional) - a sequence $\mathrm{T}_a = (a_1, a_2, \dots , a_n) \seq \R$. 
    \item Gyroscopic measurements (optional) - a sequence $\mathrm{T}_g = (g_1,g_2,\dots, g_n) \seq \R^c$.
    \item Elevation data (optional) - $\mathrm{T}_z = (z_1,z_2,\dots,z_n) \seq \R$. 
    \item other features (optional) - $\mathrm{T}_o = (o_1, o_2, \dots o_n) \seq \R^q$. These may include signature curves (in $\R^2$), other geometric features, and features learned from machine learning, all of which shall be discussed later. 
\end{enumerate}
\end{definition}

\begin{remark}
It is worth noting that the speed, acceleration, and gyroscopic sensors may not sample at the same rate as the GPS sensor - usually at a higher sampling frequency. However, g-RIPS Sendai 2022~\cite{gripssendai2022} implemented an interpolation algorithm to align/fuse these data to the GPS points. As a consequence, we do not need to worry about misaligned sampling frequencies, and we are immeasurably grateful for this foundation to work off of.
\end{remark}

We now define the map-matching problem on which we shall work for the rest of the summer:

\begin{prob}\label{prob:onlineoffline} Map-Matching Problem. Given a road network $(V, E)$ and a GPS trajectory $\mathrm{T}$ with features $F$ and coordinates $P$, match $\mathrm{T}$ to the ground truth (or closest) route taken in $(V, E)$. There are two main subsets of map matching: 

\begin{enumerate}
\item Online map matching, or "live" matching - this is intended to happen in real-time as an entity traverses a route. This has many applications for autonomous navigation systems.
\item Offline map matching - this happens after an entity's entire course of movement has been completed, used to register the complete trajectory from point A to point B. This has many applications to map synchronization and map fusion problems. 
\end{enumerate}
\end{prob}

The main goal of this paper is to develop innovative new methodologies ideally with greater generalizability, reduced error, and applicability to modern industrial applications at scale. 




\section{Background} \label{sec:background}
In this section, we briefly review several existing map matching techniques: Topological Map Matching Algorithms, Fuzzy Logic Map Matching Algorithms, Dijkstra’s Algorithm for Offline Matching, and Deep learning for Trajectory Registration.

\subsection{Topological Map Matching Algorithm}\label{topologicalmapmatching}
The classical geometric map matching algorithm, such as point-to-point, point-to-curve, and curve-to-curve map matching algorithms, uses only geometric information, i.e., coordinates of trajectory points and distance. It focuses only on the arc shape and does not reflect other information. Therefore, although it has the advantage of being simple and fast, it lacks accuracy. In order to improve the geometric map matching algorithm, the topological map matching algorithm uses other factors such as connectivity, proximity, and contiguity of edges in addition to the geometric information. Here, we briefly present Greenfeld's algorithm~\cite{Greenfeld2002MATCHINGGO}.

His algorithm consists of two parts, which he called InitialMapping() and Map(). The first algorithm InitialMapping() finds an initial matching of a point to a vertex. It is applied in the following situations:

\begin{enumerate}
    \item When the first trajectory point $p_0$ is obtained. 
    \item When the distance between the new trajectory point $p_t$ and the previous point $p_{t-1}$ exceeds a pre-selected threshold. 
    \item When the second algorithm Map() cannot continue the current map matching.
\end{enumerate}
When these situations are met, InitialMapping() algorithm is executed according to the following steps:

\begin{enumerate}
    \item Pick the closest vertex $v_0$ to a point $p_0$.
    \item Determine all the edges in the road network that are connected to $v_0$.
    \item For the next point $p_1$, pick an edge $e_1$ from the selected edges.
\end{enumerate}
After InitalMapping(), we execute the second algorithm Map(). It is the main algorithm of his approach that uses topological reasoning and a weighting method to find the correct edge $e_t$ for a point $p_t$ in question. It takes the following steps:
\begin{enumerate}
    \item Obtain the next trajectory point $p_t$.\label{greenfeldstep1}
    \item Form a line segment $p_{t-1}p_t$ by connecting $p_{t-1}$ and $p_t$.
    \item Determine whether $p_t$ still matches the previously selected edge $e_{t-1}$ by evaluating the proximity and orientation of this line segment to the edge $e_{t-1}$.\label{greenfeldstep3}
    \item If $p_t$ matches $e_{t-1}$, set $e_t \coloneqq e_{t-1}$. Otherwise, find the correct edge, $e_t$, among those connected to $e_{t-1}$ or located downstream from $e_{t-1}$ based on the same evaluation method as in the previous step.
    \item Go back to Step~\ref{greenfeldstep1}.
\end{enumerate}
The evaluation in Step~\ref{greenfeldstep3} assigns a weight to the edge $e_{t-1}$ based on the direction of the line segment $p_{t-1}p_t$, the distance between $p_t$ and $e_{t-1}$, and the intersection of $p_{t-1}p_t$ and $e_{t-1}$. This is the topological part of his algorithm.

\begin{remark}\label{rem:greenfeld}
It is worth noting that he focused on online map matching, so when considering a point at a certain time, his algorithm (basically) used no more information about future points than that point.
\end{remark}

\begin{figure}
  \centering
  \includegraphics[width=.8\linewidth]{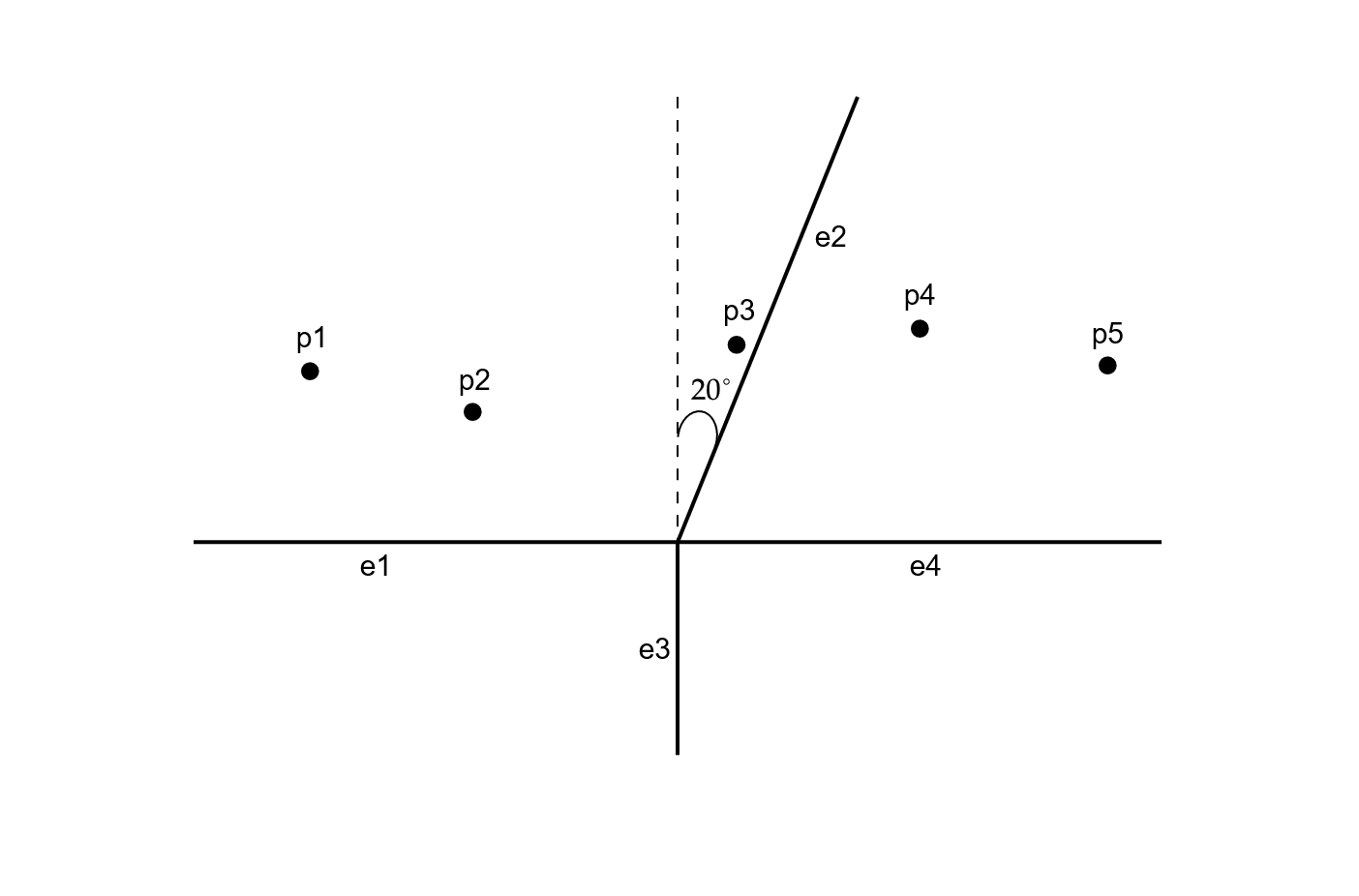}
  \caption{Mismatching in Greenfeld's algorithm}
  \label{fig:greenfeldmismatching}
\end{figure}

His program is simple and straightforward but has some problems including:
\begin{enumerate}
    \item InitialMapping() is less accurate since it simply takes the closest vertex.
    \item The proximity is regarded as the most important factor in determining weights, but this is not always the case. For example, in \cite{quddus2003general}, the following case is reported (see Figure~\ref{fig:greenfeldmismatching}): The algorithm can detect that both points $p_1$ and $p_2$ match the edge $e_1$. However, for a point $p_3$, it gives the mismatching to the edge $e_2$ instead of $e_4$.
    \item Since the algorithm only uses information available from coordinates, it is heavily influenced by outliers.
\end{enumerate}

In Subsection~\ref{subsection:ahpmapmatching}, inspired by his algorithm and also that by \cite{velaga2009developing}, we introduce a novel topological map matching algorithm that incorporates the analytic hierarchy process. It uses direction and speed data of trajectory points in addition to distance data. Moreover, in the algorithm, we use multiple weight coefficients (parameters) for each factor, depending on the map environment around the point under consideration.

\subsection{Fuzzy Inference System}\label{Fuzzy}

Fuzzy logic is an approach to computing based on degree of truth, usually represented as probability ranges from 0 to 1 rather than the usual boolean logic. Fuzzy logic aims at modeling the imprecise modes of reasoning that play an essential role of human rational decision making and can be viewed as the extension of multi-valued logic (\cite{zadeh1988fuzzy}). A Fuzzy Inference System is the framework for applying fuzzy logic to obtain a numerical output from a given input.  

\begin{figure}
    \centering
    \includegraphics[width=.8\linewidth]{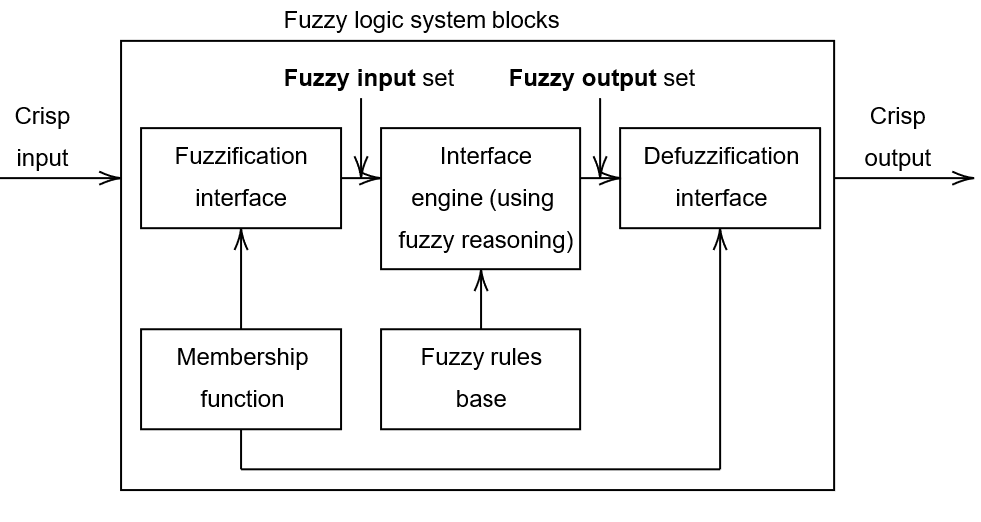}
    \caption{Fuzzification-Defuzzification pipeline.}
    \label{fig:ppl}
\end{figure}

\begin{definition}[Fuzzy Inference System] Fuzzy Inference System (illustrated in figure \ref{fig:ppl}) is a process of formulating mapping from an input to an output using some if-then rules. The FIS rule base is made of N rules of the following form : 
\begin{align*}
R_i : &\textbf{If } S_1 \text{ is } L_1^i \textbf{ and } \cdots  S_p \text{ is } L_p^i\\
      & \textbf{Then } Y_i
\end{align*}
where : 
    \begin{itemize}
        \item \textbf{Rule} $R = (R_1, R_2, \dots, R_N)$ , $R_i$ is the $i^{th}$ rule base.
        \item \textbf{Crisp input} is a vector $S = (S_1, S_2, \dots, S_p) \in \mathbb{R}^p$  
        \item \textbf{Fuzzy input}  is a vector $L = (L_1, L_2, \dots, L_N), L_j = (L_j^1, L_j^2, \dots, L_j^p) \in [0,1]^p$ such that $L_j^i \in [0,1]$, in where decision making process occurs. where $L_j^i$ represent the transformed value of input $S_j$ in rule $R_i$.
        \item \textbf{Membership function} $\mu_{L_j^i}:  \mathbb{R} \to [0,1]$ (resp. $\mu_{O_j^i})$ is a function that transformed input ($S_j$) (resp. fuzzy output $Y_j$) into fuzzy input $L_j^i$. (resp. output $O_j^i$). 
        \item \textbf{Fuzzy output} $Y = (Y_1, Y_2, \cdots, Y_q)$ such that $Y_i$ is the $i^{th}$ solution calculated from fuzzy input. 
        \item \textbf{Crisp Output} or \textbf{conclusion} $O = (O_1^1, O_1^2, \cdots ,O_q^N)$ is the linguistic output variable $Y_j$ in rule $R_i$
    \end{itemize}
\end{definition}

Fuzzy inference system algorithm consists of three main steps illustrated by figure \ref{fig:ppl} : 

\begin{enumerate}
    \item Fuzzification: fuzzifying input values ($S$) with membership functions($\mu_{L_j^i}$) to a fuzzy input($L$).
    \item Fuzzy inference step: operating all applicable rules from fuzzy input ($L$) to fuzzy output ($Y$).
    \item Defuzzication: De-fuzzifying fuzzy output($Y$) set back to a crisp output value ($O$).
\end{enumerate}

In the following example we will discuss a simplified version Fuzzy Inference System given by Quddus~\cite{quddus2006high}. 

\begin{example}\label{ex:FIS}
\begin{figure}[H]
\centering
    \includegraphics[width=.6\linewidth]{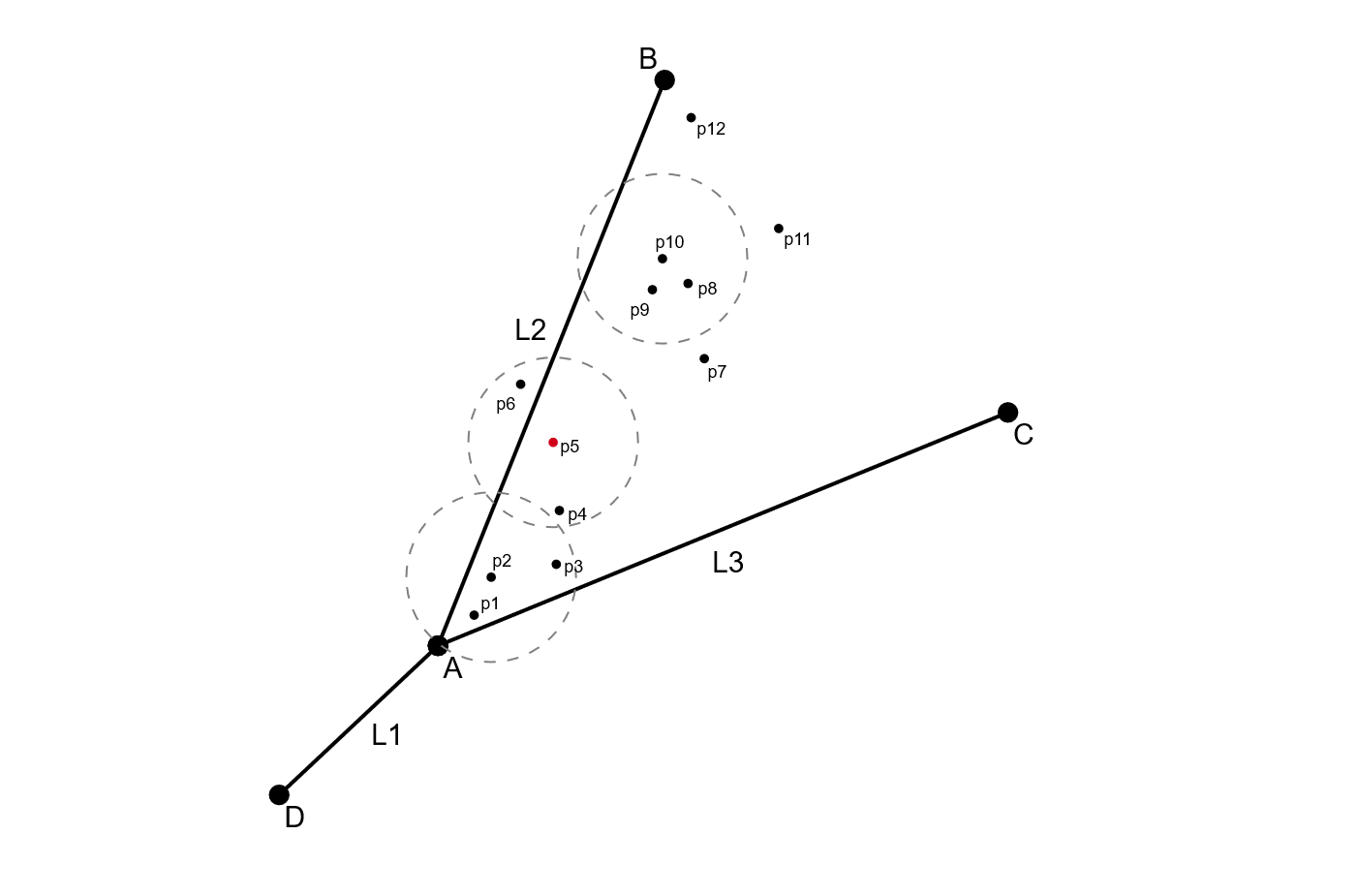}
    \caption{three legged junction example}
    \label{jnct}
\end{figure}

Suppose we would like to decide which link (DA, AB, or AC) does vehicle at position p2 travels in a three legged junction illustrated in figure \ref{jnct}. Denote PQ as the perpendicular distance from P to link AB, $\theta$ as the direction of the vehicle at P obtained from the navigation sensor, and $\Delta\varphi = | 90^\circ - \theta|$ as the angular difference of the vehicle. Suppose PQ and $\Delta\varphi$ are the two primary determining factors, also referred to as crisp input,  as to whether P is matched to AB. A knowledge based rules can be formed based on these two inputs, for example : 

\begin{enumerate}
    \item $R_1$ : If PQ is short and $\Delta\varphi$ is small then the possibility of matching P on link AB is high.
    \item $R_2$ : If PQ is long  and $\Delta\varphi$ is large then the possibility of matching P on link AB is low. 
\end{enumerate}

Where terms such as short (or long) and small (or large) are linguistic values that can be defined on the range of PQ and $\Delta\varphi$. We can start step 1 by transforming our crisp input PQ ($S_1$) and $\Delta\varphi (S_2)$ into a fuzzy input denoted by $L_1 = (L_1^1, L_1^2)$ and $L_2 = (L_2^1, L_2^2)$ using some membership function ($\mu_{L_{j}^i}$) where $i, j \in {1,2}$ that is known from domain knowledge.

Once the input is fuzzified, we can proceed to the next step by evaluating a fuzzy output using some fuzzy operator. The "if" part of the rule (e.g. PQ is short) is called an antecedent or premises. When the rule consists of more than one antecedent we can use fuzzy operators to evaluate the results of the rule, also commonly called rule strength. Two fuzzy operators that are normally used are the AND and OR operator. Min (minimum) and prod (product) are popular functions for AND operator, while max (maximum) and probabilistic OR are widely used for OR operator \cite{quddus2007current}.

In this example, suppose  PQ is equal to 15m and $\varphi = 15^\circ$. Figure \ref{fig:calc1} and figure \ref{fig:calc2} illustrate how we can calculate the fuzzy inputs ($L_1, L_2)$ and strength of each rules. The strength of each rules are calculated by applying the min function for the \textbf{AND} operator. Since the fuzzy input from the first rule are $L_1^1 =0.5$ and $L_2^1 = 0.8$, the strength of the first rule is equal to $min(L_1^1, L_2^1) = 0.5$. Similarly, since the fuzzy input for rule two is equal to  $L_1^2 = 0.4$ and $L_2^2 = 0$) so the strength of the first rule is equal to $min(L_1^2 ,L_2^2) = 0$.   

\begin{figure}[H]
  \begin{minipage}[t]{0.3\linewidth}
    \centering
    \includegraphics[width=1\linewidth]{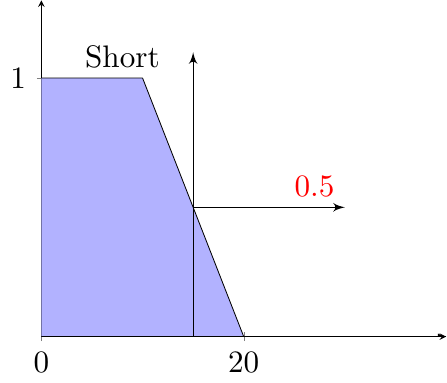}
    \caption{}
    \label{fig:l11}
    \vspace{4ex}
  \end{minipage}
  \begin{minipage}[t]{0.3\linewidth}
    \centering
    \includegraphics[width=1\linewidth]{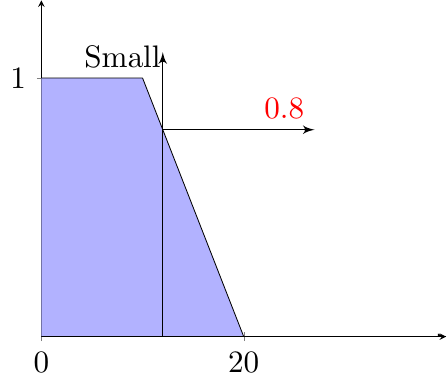}
    \caption{}
    \label{fig:l21}
    \vspace{4ex}
  \end{minipage} 
  \begin{minipage}[t]{0.3\linewidth}
    \centering
    \includegraphics[width=1\linewidth]{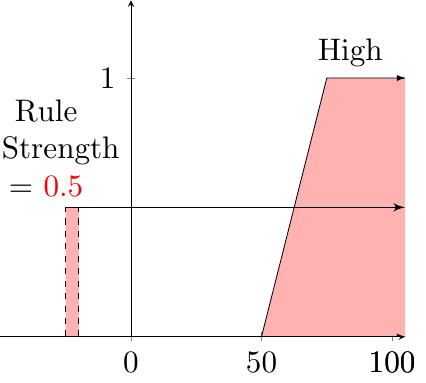}
    \caption{} 
    \label{fig:rs1}
    \vspace{4ex}
  \end{minipage}
  \caption{Rule 1}
  \label{fig:calc1}
\end{figure}


\begin{figure}[H] 
  \begin{minipage}[t]{0.33\linewidth}
    \centering
    \includegraphics[width=1\linewidth]{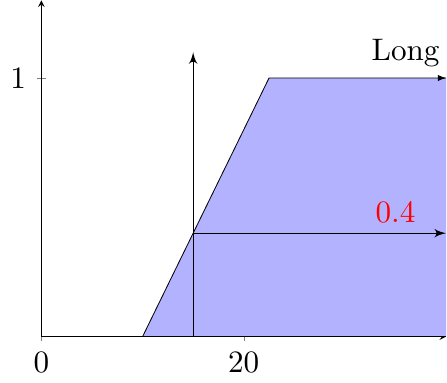}
    \caption{} 
    \label{fig:l12}
    \vspace{4ex}
  \end{minipage}%
  \begin{minipage}[t]{0.33\linewidth}
    \centering
    \includegraphics[width=1\linewidth]{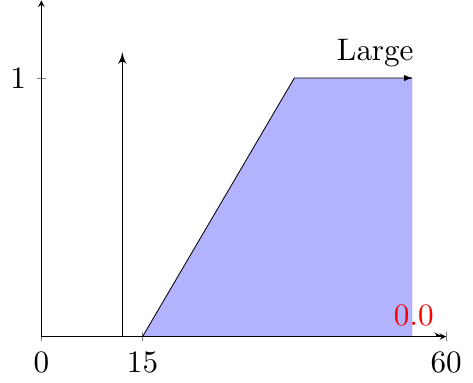}
    \caption{} 
    \label{fig:l22}
    \vspace{4ex}
  \end{minipage}%
  \begin{minipage}[t]{0.33\linewidth}
    \centering
    \includegraphics[width=1\linewidth]{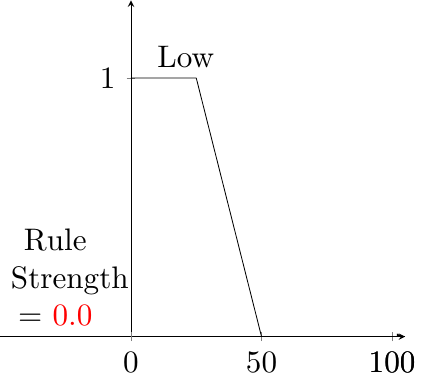}
    \caption{}
    \label{fig:rs2}
    \vspace{4ex}
  \end{minipage} 
  \caption{Rule 2}
  \label{fig:calc2}
\end{figure}

We can then convert this rule strength into a fuzzy output($Y = Y_1, Y_2$) using two membership function shown in the figure \ref{fig:rs1} and \ref{fig:rs2}. Lastly we can proceed to defuzzifying step where we aggregate these fuzzy output into a crisp output($O_1$). Area of the centroid (Mandani FIS method) or linear combination of each output(Sugeno FIS method) are one of the few common de-fuzzifying methods that can be used for our final decision making process. 
\end{example}

\subsection{Dijkstra's Algorithm for Offline Matching}

Rather than do an exhaustive brute-force map matching algorithm, which requires $\mathcal{O}(n^2)$ computations, many state of the art algorithms (e.g. Valhalla \cite{valhalla}) employ Dijkstra's algorithm \cite{dijk}, and \cite{gripssendai2022} utilized Dijkstra's \cite{dijk} as well. For a given graph, this employs a greedy registration framework with edge weights to identify the path from a starting point to end point that best minimizes the total edge weights. Interestingly, Dijkstra formulated the algorithm in only twenty minutes and published it as a 3 page "note on two problems in connexion with graphs", and it remains one of the most infleuntial algorithms in graph theory and shortest path methods.

The algorithm is shown in Algorithm \ref{dijk_alg}. Select $p_0$ as the starting node and, let $d_i$ denote the "distance" (i.e. lowest sum of weights on connecting edges) from $p_0$ to node $p_i$.

\begin{algorithm}

\KwIn{Graph $G$, source node $p_0$}
\KwOut{Shortest distances from $p_0$ to all other nodes}
Initialize distance to $p_0$ as 0 and all other nodes to $\infty$
Create a priority queue $Q$;
Insert $p_0$ into $Q$.

\While{$Q$ is not empty}{
Extract node $u$ with minimum distance from $Q$;
\ForEach{neighbor $v$ of $u$}{
\If{distance to $v$ through $u$ is shorter than current distance}{
Update distance of $v$ to the new shorter distance;
}
\If{$v$ is not visited}{
Add $v$ to $Q$;
}
}
}

\caption{Dijkstra's Algorithm}
\label{dijk_alg}
\end{algorithm}

Note: this algorithm was typeset by ChatGPT \cite{chatgpt}, with minor modifications, corrections, and changes of notation. We have verified with \cite{dijk} that it is indeed correct. This is the only portion of the report written or typeset using ChatGPT.

\subsection{Deep learning for Trajectory Registration}

Relatively little has been done (publicly) in the way of deep learning for GPS trajectory registration due to the lack of large, publicly available labelled datasets with many features. \cite{ml4speed} utilized a rather exhaustive geometric, topological, and speed matching (i.e. max feasible speed of the roadway) analysis to propose candidate routes. Finding the rigid speed matching unideal, they deploy machine learning to predict each road's speed using a bidirectional Conv-LSTM RNN. The training of this Conv-LSTM requires road-wise traffic usage data, which we do not have access to. Other approaches such as \cite{sun_hmm_rl} also employ historical data to train a Hidden Markov Model (HMM), but their data is not public either. \cite{ml4repositioning} employed a simple 2 layer neural network to reposition the GPS points slightly prior to inputting them into the registration algorithm - interestingly, they implement it for real time applications, which were trained using the horizontal displacement between the GPS point and the ground truth trajectory for selections from the OpenStreetMaps \cite{openstreetmap} traces data.

\section{Our Preliminary Approach}
\label{sec:approach}

\subsection{Stay Point Mitigation and Outlier Detection by DBSCAN}
It is known that stay points and outliers which may be caused by GPS errors may prevent map matching algorithms from finding the correct route.
It is researched in ~\cite{jafarlou_improving_nodate} that Density-Based Spatial Clustering of Applications with Noise (DBSCAN)~\cite{ester1996density} may be utilized to detect and mitigate stay points in a GPS trajectory.
In this research, we developed a method to utilize DBSCAN to detect outliers.
Furthermore, in addition to these applications, we propose a method to automatically determine a parameter passed into the algorithm according to each input of points.
Although we could not actually integrate the methods that we developed into map matching algorithms due to the time constraints of the research project, we managed to implement the methods in Python and expect that it will improve the performances of our algorithms.
The details of these methods are talked about in Subsection ~\ref{subsection:dbscan}.

\subsection{Datasets}

The GPS data that is publicly available is severely lacking, especially for the purposes of deep learning. State of the art open-source frameworks like Valhalla \cite{valhalla} utilize over 18 million verified trajectories to train their registration framework (a Hidden Markov Model), but this trajectory data is not open-source. Companies like Google and Apple, the creators of two of the most popular navigational smartphone applications, certainly have access to far more trajectory data than that to train their models.

If we are to come anywhere close to the success of existing industrial models with a data-driven approach, we are going to need an extensive amount of data. The largest datasets we can access (or plan to assemble) are thus:

\begin{enumerate}
    \item KCMMN - "Dataset for testing and training map-matching methods" \cite{KCMMNdataset} - 100 different trajectories spanning over 5,000 km of roads across the globe. Features are restricted to GPS coordinates and timestamps only; has verified ground truth trajectories. 
    \item BDD100K \cite{yuBDD100KDiverseDriving2020} - 100,000 trajectories of 40 second travel sequences. Contains GPS coordinates, timestamps, IMU data (highly accurate speed, direction, and gyroscopic data), and videos; has no ground truths. 
    \item OpenStreetMaps Traces \cite{openstreetmap} - very large, publically available dataset consisting of GPS coordinates and elevation data for many trajectories collected around the globe. The dataset is constantly growing. Notably there is no IMU data or ground truth. 
    \item 2022 G-RIPS Mitsubishi-A data - collected from former participants walking around Sendai. Unclear if it is just GPS coordinates and timestamps or if it contains IMU data as well. Should have ground truths, but we have not yet seen the data at this time.
\end{enumerate}

We can summarize the publicly available data as follows and see the clear gaps:\\\\
\begin{figure}[H]
\centering
\begin{tabular}{ |p{3.5cm}||p{1.7cm}|p{1.7cm}|p{1.8cm}| p{2cm}| p{2cm}|}
 \hline
 
Dataset & Raw GPS Timeseries & IMU data & Velocity & Elevation & Ground Truth \\
 \hline
  KCMMN \cite{KCMMNdataset}   & V    & X &   X & X & V\\
 \hline
 BDD100K \cite{yuBDD100KDiverseDriving2020} &  V  & V   & V & X & X \\
 \hline 
 OpenStreetMaps \cite{openstreetmap} & V& X & V & V & X\\
 \hline
 EnviroCar \cite{EnviroCar} & V & X & V & X & X \\
 \hline
\end{tabular}
\end{figure}
This incentivizes some means to infer IMU/Velocity/Elevation data for the KCMMN dataset, the only one with ground truth: i.e. {\em data fusion}.

\subsection{Data Fusion: Estimating IMU Data for KCMMN}
In the absence of a large trajectory dataset equipped with both IMU data and ground truths, we propose to first leverage the BDD100K data (with IMU data) to estimate IMU data for the KCMMN trajectories. This can be readily achieved using a physics-informed neural network with filtering, or a relatively lightweight GNN, RNN, or transformer. Such data may also be estimated for the OpenStreetMaps trajectory data, provided it works well for KCMMN.  

Speed and direction of travel are of greater intuitive interest than the gyroscopic measurements, but we should try to estimate all possibly useful values that we can. These quantities may also be approximated directly from the GPS trajectories and timestamps with some form of local filtering and/or regression.

\subsubsection{Preliminary Results and Changes to Approach}

We first implemented a 3-layer NNConv GNN using PyTorch Geometric (PyG) to model the IMU data using graphs built off the GPS trajectories using the BDD100K dataset; each layer of which used a 3-layer network to weight the features based on the differences in the input trajectories. Overall, this failed, despite modifications to the loss to use percent error instead of absolute error. Despite PyG having a DataParallel module (notably one with exceptionally poor documentation and no examples), training with multiple GPU's was exceptionally slow and tedious (it was only realized later that CSV's read in 16x slower than NPZ files, but even after converting the data, it was still very slow). It remains a very curious matter. Finally, after PyG's distributed processes crashed two A-100 GPU's at the Minnesota Super-computing Institute, we resorted to abandon PyG and try a 1-dimensional CNN instead.

It appears that the 1D CNN experienced vanishing gradients, ultimately returning all 0's. It was then thought that this was due to the padding (most trajectories in the BDD100K had about 40 points, but a few had 80), but after implementing a masking framework for the loss, this still converged to 0's. After various attempts to remedy this, we ultimately decided to focus on training the edge affinity function instead.

Given the difficulties in training an accurate model with the BDD100K dataset \cite{yuBDD100KDiverseDriving2020} (and paired with unanswered requests for clarification on the units on the BDD100K data), we elected to approximate the velocities and directions of travel for the KCMMN dataset \cite{KCMMNdataset} using simple 1D differentiation filters. While most of the data follows regular sampling, a few points have irregular timesteps, which makes higher order differentiation filters a little more challenging to implement. Ultimately we used the neighboring two point approximation to approximate the velocity:
$$v(x[i]) \approx \frac{\overrightharp{p}[i+1] - \overrightharp{p}[i-1]}{t[i+1] - t[i-1]}$$
From this, we approximate the speed $\no{v(\ar{p}[i])}$ and direction of travel as the angle from the $x$-axis. This allowed our Fuzzy-Inference and AHP based methods to run on KCMMN. In the future, far more robust approximations should be used, i.e. local parametric curve fitting from the timestamps, but time did not allow this to be implemented.

\section{Map Matching Algorithms}\label{sec:map-matching-algorithms}

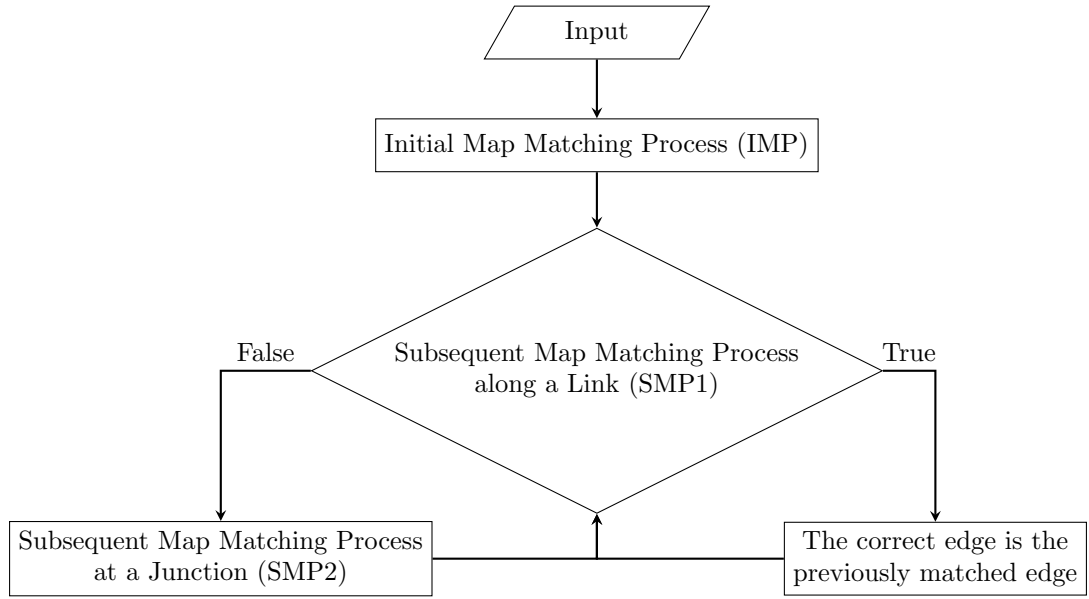
\begin{figure}
    \centering
    \begin{tikzpicture}
        \node[io] at (4.5, 3) (input) {Input};
        \node[process] at (4.5,1.5) (IMM) {Initial Map Matching Process (IMP)};       
        \node[decision, align=center, yshift=-0.5cm] at (4.5,-1) (check) {Subsequent Map Matching Process\\
        along a Link (SMP1)};
        \node[process, align=center, yshift=-1cm] at (9,-3) (asbefore) {The correct edge is the\\ previously matched edge};
        \node[process, align=center, xshift=-0.5cm, yshift=-1cm] at (0,-3) (MMJ) {Subsequent Map Matching Process\\
        at a Junction (SMP2)};
        \draw[arrow] (input)--(IMM);
        \draw[arrow] (IMM)--(check);
        \draw[arrow] (check)-|(asbefore)node[above,pos=0.25]{True};
        \draw[arrow] (check)-|(MMJ)node[above,pos=0.25]{False};
        \draw[arrow] (MMJ)-|(check);
        \draw[arrow] (asbefore)-|(check);
    \end{tikzpicture}
    \caption{The flowchart of our map matching algorithms}
    \label{fig:flowofmapmathing}
\end{figure}

In this section we propose two map matching algorithms: AHP map matching algorithm and Fuzzy logic map matching. Our algorithms are classified as online map matching (Problem~\ref{prob:onlineoffline}). They find the correct edge for each trajectory point. The flowchart of the algorithms is shown in Figure~\ref{fig:flowofmapmathing}. Both algorithms take similar steps, which are: initial map matching process (IMP), subsequent map matching process along a link (SMP1), and subsequent map matching process at a junction (SMP2). Once input is obtained, IMP is executed to find the correct edge for the first (few) trajectory point(s). After IMP, SMP1 is performed to verify whether the next point still matches the same edge as the previous point. If yes, the algorithms repeat SMP1 for the next point. If not, they execute SMP2 to find the correct edge for that point. Then they go back to SMP1 and continue this process until they reach the last trajectory point. The details of each algorithm are explained in the following Subsections~\ref{subsection:ahpmapmatching},~\ref{subsection:fuzzymm}.

\subsection{AHP Map Matching Algorithm}\label{subsection:ahpmapmatching}

The AHP stands for the analytic hierarchy process. This is a decision-making method that combines mathematical analysis with human judgment. It utilizes hierarchical classification to deal with complex and abstract information. AHP has not been used much for map matching. To the best of our knowledge, the only paper that mentions AHP in the context of map matching is the one by~\cite{mahpour2022improvement}, but their method differs from ours. Therefore this algorithm can be considered relatively new and is also simple and easy to understand.

\begin{figure}
    \centering
    \begin{tikzpicture}
        \node[process] at (0,0) (aim) {Choose the correct edge};
        \node[process, below of=aim, yshift=-0.3cm] (direction) {Direction};
        \node[process, left of=direction, xshift=-3cm] (distance) {Distance};
        \node[process, right of=direction, xshift=3cm] (turn) {Turn restriction};
        \node[process,below of=distance, yshift=-0.3cm, xshift=1cm] (candidate1) {Candidate edge 1};
        \node[below of=direction, yshift=-0.3cm, xshift=0.1cm] (dots) {$\cdots$};
        \node[process,below of=turn, yshift=-0.3cm, xshift=-1cm] (candidaten) {Candidate edge $n$};
        \draw[thick] (aim)--(direction);
        \draw[thick] (aim)--(distance);
        \draw[thick] (aim)--(turn);
        \draw[thick] (distance)--(candidate1);
        \draw[thick] (distance)--(candidaten);
        \draw[thick] (direction)--(candidate1);
        \draw[thick] (direction)--(candidaten);
        \draw[thick] (turn)--(candidate1);
        \draw[thick] (turn)--(candidaten);
    \end{tikzpicture}
    \caption{AHP layer}
    \label{fig:ahplayer}
\end{figure}
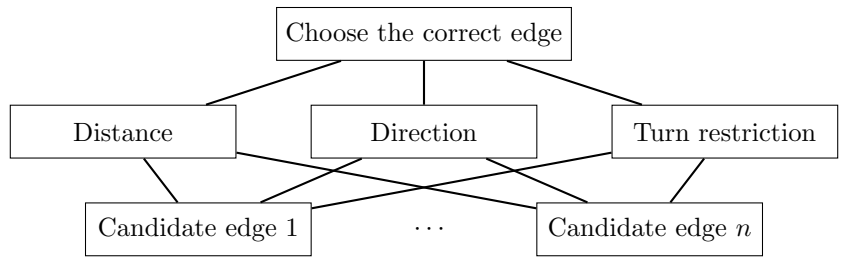

The AHP is incorporated into IMP and SMP2 parts (Figure~\ref{fig:ahplayer}). As input, we use a road network, trajectory points, and the speed and direction data of each trajectory point.

\subsubsection{Initial Map Matching Process (IMP)}
The purpose of the initial map matching process (IMP) is to specify the first matching. The IMP takes the following steps:
\begin{enumerate}
    \item Identify a set of candidate edges.
    \item Assign a weight to each candidate edge using AHP based on distance and direction data.
    \item Take the highest weight edge as the correct edge for that point.
\end{enumerate}
In the first stage, we check whether the speed of the vehicle at the first point $p_0$ is less than 3m/s or not. If yes, we skip the analysis of $p_0$ and run IMP for the next point $p_1$. This is because if the vehicle speed is less than 3m/s, the GPS data is less reliable \cite{taylor2001road,ochieng2003map}. We continue this speed check until the first point where the speed is more than $\mathrm{3m/s}$ is obtained. Once such a point, say $p_t$, is obtained, next we draw the error polygon and take edges that intersect or are contained in this polygon as candidate edges. So candidate edges are the edges that have the possibility of matching $p_t$. Let $e_1,\ldots,e_n$ be the candidate edges. After identifying the candidate edge, we define a weight for each candidate edge using AHP. A weight is given based on distance data and direction data of each candidate edge. To do this, we first consider weights $w_1^{\mathrm{dist}},\ldots,w_n^{\mathrm{dist}}$ for distance. To begin we construct the pairwise comparison matrix $M_{\mathrm{dist}}=[\alpha_{ij}]$ for distance. Each $(i,j)$-component of $M_{\mathrm{dist}}$ is defined by Table~\ref{pairmatdist}. Here, the distance $\operatorname{dist}(p,e)$ between a point $p$ and an edge $e$ is defined by 
\begin{equation*}
    \operatorname{dist}(p,e) \coloneqq \sup_{x\in e} \operatorname{d}(p,x),
\end{equation*}
where $\operatorname{d}(p,x)$ is the usual euclidean distance between two points (Figure~\ref{defofdist}). Then we take the geometric mean $g_i$ of each row and let $S=g_1+\cdots+g_n$. Finally the weight $w_i^{\mathrm{dist}}$ for distance of the candidate edge $e_i$ is determined by $e_i=g_i/S$. We repeat the same procedure to obtain weights $w_1^{\mathrm{dir}},\ldots,w_n^{\mathrm{dir}}$ for distance using the hierarchical classification of Table~\ref{pairmatdir}. Now each candidate edge has two weights: one for distance and another for direction. Finally the total weight $\operatorname{TW}(e_i)$ for $e_i$ is defined by
\begin{equation*}
    \operatorname{TW}(e_i) \coloneqq c^{\mathrm{dist}}w_i^{\mathrm{dist}} + c^{\mathrm{dir}}w_i^{\mathrm{dir}},
\end{equation*}
where $c^{\mathrm{dist}}$ and $c^{\mathrm{dir}}$ are the coefficients that reflect the relative importance of the factors. These values are provided by \cite{velaga2012improving} and given by Table~\ref{relativeimportance}, which vary depending on the map environment (see also subsection~\ref{mapenvironment}). After this process we select the highest weight edge as the correct edge for $p_t$.

\begin{table}
    \centering
        \centering
        \caption{The pairwise comparison matrix for distance. $d_i\coloneqq\operatorname{dist}(p_t,d_i)$}
        \label{pairmatdist}
        \begin{tabular}{c|c}
            \hline
            $\alpha_{ij}$ & range \\
            \hline\hline
            1 & $0\leq d_j-d_i\leq1$\\
            2 & $1< d_j-d_i\leq3$\\
            3 & $3< d_j-d_i\leq5$\\
            4 & $5< d_j-d_i\leq7$\\
            5 & $7< d_j-d_i\leq9$\\
            6 & $9< d_j-d_i\leq11$\\
            7 & $11< d_j-d_i\leq13$\\
            8 & $13< d_j-d_i\leq15$\\
            9 & $15< d_j-d_i$\\
            $1/\alpha_{ji}$ & $d_j-d_i<0$   
        \end{tabular}
\end{table}

\begin{table}
        \centering
        \caption{The pairwise comparison matrix for direction. $\theta_i$ is the angle difference between the direction of $p_t$ and the direction of $e_i$}
        \label{pairmatdir}
        \begin{tabular}{c|c}
            \hline
            $\beta_{ij}$ & range \\
            \hline\hline
            1 & $0\leq \theta_j-\theta_i\leq10$\\
            2 & $10< \theta_j-\theta_i\leq30$\\
            3 & $30< \theta_j-\theta_i\leq50$\\
            4 & $50< \theta_j-\theta_i\leq70$\\
            5 & $70< \theta_j-\theta_i\leq90$\\
            6 & $90< \theta_j-\theta_i\leq110$\\
            7 & $110< \theta_j-\theta_i\leq130$\\
            8 & $130< \theta_j-\theta_i\leq150$\\
            9 & $150< \theta_j-\theta_i$\\
            $1/\beta_{ji}$ & $\theta_j-\theta_i<0$   
        \end{tabular}
\end{table}

\begin{figure}
    \centering
    \begin{tikzpicture}
        \draw[ultra thick] (0,0)--(2,0) node[below,pos=0.5]{$e$};
        \draw[ultra thick] (3,0)--(5,0) node[below,pos=0.5]{$e$};
        \draw[dashed, thick] (5,0)--(5.5,0)--(5.5,1.5);
        \draw[very thick, red] (5,0)--(5.5,1.5) node[left,pos=0.5]{$d$};
        \draw[very thick, red] (1,0)--(1,1.5) node[left,pos=0.5]{$d$};
        \draw (1,0.2)--(1.2,0.2)--(1.2,0);
        \draw (5.3,0)--(5.3,0.2)--(5.5,0.2);
        \filldraw (1,1.5) node [left] {$p$} circle [radius=1pt];
        \filldraw (5.5,1.5) node [left] {$p$} circle [radius=1pt];
    \end{tikzpicture}
    \caption{Distance $\operatorname{dist}(p,e)$ between a point and an edge}
    \label{defofdist}
\end{figure}
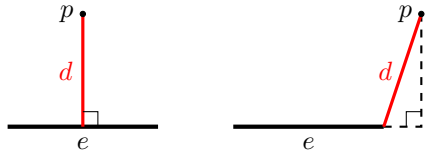
\begin{table}
    \centering
    \caption{Relative importance}
    \label{relativeimportance}
    \begin{tabular}{c|c|c|c}
        \hline
        & urban & suburban & rural\\
        \hline\hline
        $c^{\mathrm{dist}}$ & 0.0806 & 0.4376 & 0.5563\\
        $c^{\mathrm{dir}}$ & 0.3715 & 0.4642 & 0.4237\\
        $c^{\mathrm{turn}}$ & 0.5479 & 0.0982 & 0.020
    \end{tabular}
\end{table}

\subsubsection{Subsequent Map Matching Process along a Link (SMP1)}
After each matching, we always run the subsequent map matching process along a link (SMP1) for the next trajectory point. In SMP1 we verify whether the next point matches the previously selected edge. Let $p_t$ be the current point, $e$ be the previously selected edge for the previous point $p_{t-1}$, and $q_{i-1}$ be the projection point of $p_{t-1}$ onto $e$. At first, if the speed of $p_t$ is zero, then we automatically conclude that $p_t$ matches $e$. In other cases, we use two factors to determine this: 
\begin{enumerate}
    \item How far $p_t$ is from the next junction point.
    \item How small the angle difference between the direction of $p_t$ and that of $p_{t-1}$ is.
\end{enumerate}
For the first part, we set $d_1$ as the distance between $q_{i-1}$ and the next junction point and $d_2$ as the product of the speed of $p_{t-1}$ and time interval between $p_{t-1}$ and $p_t$. We then define $\Delta d \coloneqq d_1-d_2$. For the second part, we calculate $\Delta h$ as the angle difference between the direction of $p_t$ and that of $p_{t-1}$. If both conditions $\Delta d>30$ and $\Delta h<5$ are satisfied, then we conclude that $p_t$ still matches the previously selected edge. Otherwise, we proceed to the subsequent map matching process at a junction.

\subsubsection{Subsequent Map Matching Process at a Junction (SMP2)}
The subsequent map matching process at a junction (SMP2) is executed only for the point that does not meet the SMP1. SMP2 takes a similar process as IMP, which is:
\begin{enumerate}
    \item Identify a set of candidate edges.
    \item Assign a weight to each candidate edge using AHP based on distance, direction, and turn restriction data.
    \item Take the highest weight edge as the correct edge for that point.
\end{enumerate}
Let $p_t$ be the current point. The first part is exactly the same as the one in IMP and we assume that $e_1,\ldots,e_n$ are candidate edges. In the second part, similar processes used in IMP are also employed except that we use a new factor, which is turn restriction data, in addition to distance and direction data. So the method for obtaining the weights for distance and direction is the same as for the IMP. The turn restriction data reflects if the vehicle on the previously selected edge can legally turn onto each candidate edge, that is if each candidate edge is connected to the previously selected edge and is not the wrong way down a one-way street. The pairwise comparison matrix $M_{\mathrm{turn}}=[\gamma_{ij}]$ for turn restriction is defined by Table~\ref{pairmatturn}, and weights $w_1^{\mathrm{turn}},\ldots,w_n^{\mathrm{turn}}$ for turn restriction are obtained the same way as the other two weights. At this stage, each candidate edge has three weights: one for distance, another for direction, and the third one for turn restriction. Then the total weight $\operatorname{TW}(e_i)$ for $e_i$ is defined by
\begin{equation*}
    \operatorname{TW}(e_i) \coloneqq c^{\mathrm{dist}}w_i^{\mathrm{dist}} + c^{\mathrm{dir}}w_i^{\mathrm{dir}} + c^{\mathrm{turn}}w_i^{\mathrm{turn}},
\end{equation*}
where the coefficients $c^{\mathrm{dist}}, c^{\mathrm{dir}}$, and $c^{\mathrm{turn}}$ are given by Table~\ref{relativeimportance}. Finally, the highest weight edge is selected as the correct edge for $p_t$.

\begin{table}
    \centering
    \caption{The pairwise comparison matrix for turn restriction}
    \label{pairmatturn}
    \begin{tabular}{c|c}
        \hline
        $\gamma_{ij}$ & range \\
        \hline\hline
        1 & If the vehicle can (or cannot) legally turn onto both $e_i$ and $e_j$\\
        9 & If the vehicle can legally turn onto $e_i$ and cannot turn onto $e_j$\\
        1/9 & If the vehicle can legally turn onto $e_j$ and cannot turn onto $e_i$
    \end{tabular}
\end{table}

\subsubsection{The Map Environment}\label{mapenvironment}
In IMP and SMP2, we need to specify the map environment for each trajectory point to determine which coefficient values should be used. As \cite{velaga2012improving} reported, the importance of each factor depends on the map environment. We determine the map environment for each trajectory point individually rather than for the entire road network because some road networks may have multiple features, such as a combination of urban and suburban areas. The calculation for finding the map environment is based on \cite{velaga2012improving}. First we draw the circle of radius 200 ($\mathrm{m}$) centered at the current point $p_t$. Next we count the number $N$ of junction points within this area and calculate the total length $L$ ($\mathrm{km}$) of roads within this area. Then the ratio $N/L$ is used to determine the map environment. If the ratio is greater than 6.81 we conclude that the map environment around $p_t$ is urban, if the ratio is smaller than 2.88, then the map environment is assumed rural. Otherwise, the map environment is considered suburban.


\subsection{Fuzzy Logic Map Matching Algorithm}\label{subsection:fuzzymm}

A Fuzzy-logic Map Matching algorithm utilizes Fuzzy Inference System to make a decision. Three different Fuzzy Inference System (FIS) is implemented for IMP, SMP 1, and SMP 2. Quddus \cite{quddus2007current} proposes using Takagi-Sugeno-Kang that averages out the rule outputs
\begin{equation*}
    Z = \dfrac{\Sigma_{i = 1}^ N \omega_i Z_i}{\Sigma_{i = 1}^ N \omega_i}
\end{equation*}
Where $\omega_i$ is the rule strength calculated from the fuzzy rules and $Z_i$ is a defuzzifcation function. Since the output of this FIS is the likelihood of matching the position fix to the candidate link. The constants used to calculate the output are 10 when output is low, 50 when output is average and 100 when output is high. 

We also introduces a set of weight $a_i$, that represents on how confident we are about the output result produced by the $i^{th}$ rule, where $\sum_{i =1}^ n a_i = 1$.  Our new proposed rule aggregation  is : 
\begin{equation*}
    Z^* = \dfrac{\Sigma_{i = 1}^ N \omega_i a_i Z_i}{\Sigma_{i = 1}^ N a_i * \omega_i}
\end{equation*}
We will discuss more about the reasoning to include this new weight in section \ref{section:FLMM_res}

\subsubsection{Initial Map Matching Process (IMP)}

The Initial Map Matching starts by identifying all possible candidate links inside an elliptical error confidence region around position fix (GPS location) based on some error model. FIS is then calculated for all the candidate link and link with the highest FIS score is then selected. IMP processed is then repeated until the matched link is picked for three consecutive position fix.  Quddus~\cite{quddus2006high} proposed using four input variables in the IMP step, which includes : 1) Speed of the vehicle, 2) Heading error, 3) perpendicular distance and 4) contribution of satellite geometry to the positioning error, which represented by the horizontal dilution of precision (HDOP). Since KCMMN data set doesn't have a HDOP value, therefore we decided not utilize HDOP as one of our input. Sigmoidal Membership function is chosen in the fuzzification. We then modified the fuzzy rules in order to accommodate our selection of input. We proposes these following rules to calculate fuzzy output in this FIS : 

\begin{figure}[H]
\centering
  \centering
  \includegraphics[width=.7\linewidth]{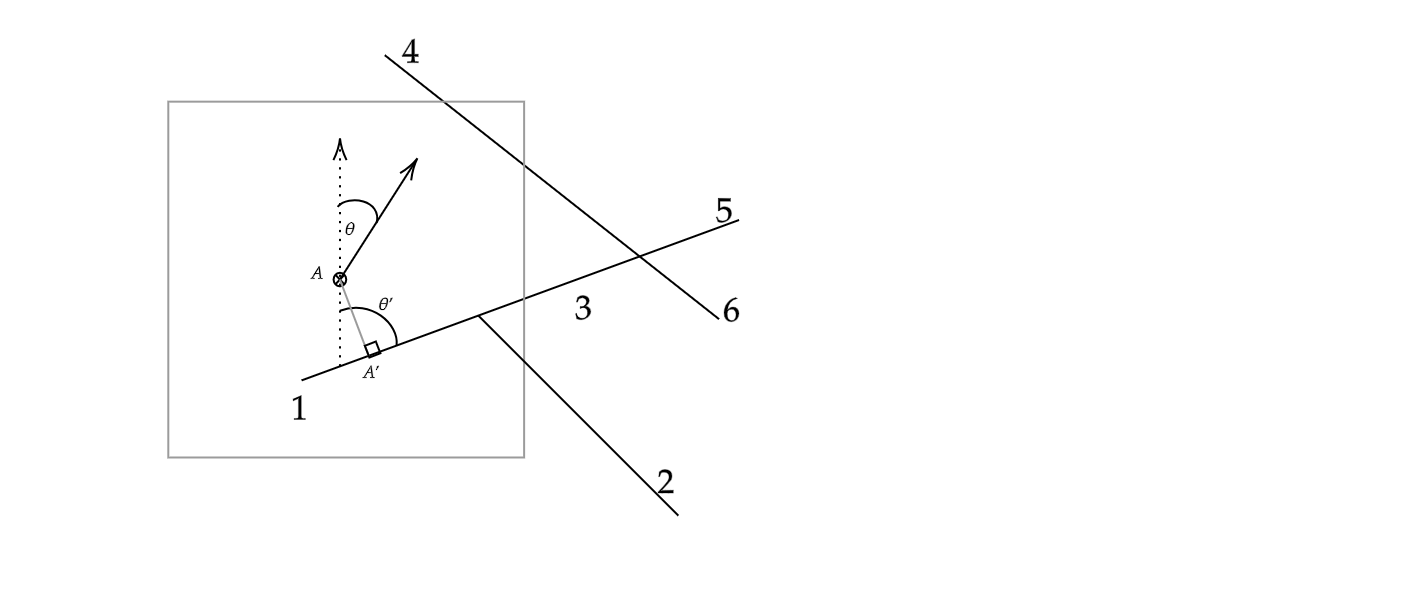}
  \vspace{-20pt}
  \caption{Four candidate link (1, 2, 3, 4) is detected in the error region. Heading error is defined as the difference between $\theta$ and $\theta$'}. 
  \label{fig:IMO}
\end{figure}

\begin{enumerate}
    \item if speed is high and heading error is small then output is average.
    \item if speed is high and heading error is small then output is low. 
    \item if perpendicular distance is short and speed is high then output is high. 
    \item if perpendicular distance is long and speed is low then output is low.
    \item if perpendicular distance is short and heading error is small then output is high.
    \item if perpendicular distance is long and heading error is large then output is low. 
\end{enumerate}

 The FIS is applied to all links within the confidence region and the link which gives the highest likelihood is taken as the correct link among the candidate link. Since the link for the first position fix may not be the actual link, IMP step can be performed to a few first position fix. If the FIS identifies the same link for those position fixes then the link is chosen as a first correct link. 

\subsubsection{Subsequent Map Matching Process along a Link (SMP1)}
Once the Initial link is chosen, SMP 1 are deployed to track whether the next position fix is still travelling through the current selected link. The input for the FIS are the speed of the vehicle, Heading Increment ($\abs{\theta - \theta'}$), $\alpha$ and $\beta$ and $\Delta d = d - d_2$, where $d_2$ speed of the previous position fixed multiplied by the time difference. In this Implementation we followed the rules proposed by Gorte \cite{gorte2014} :

\begin{figure}[H]
\centering
  \centering
  \includegraphics[width=.7\linewidth]{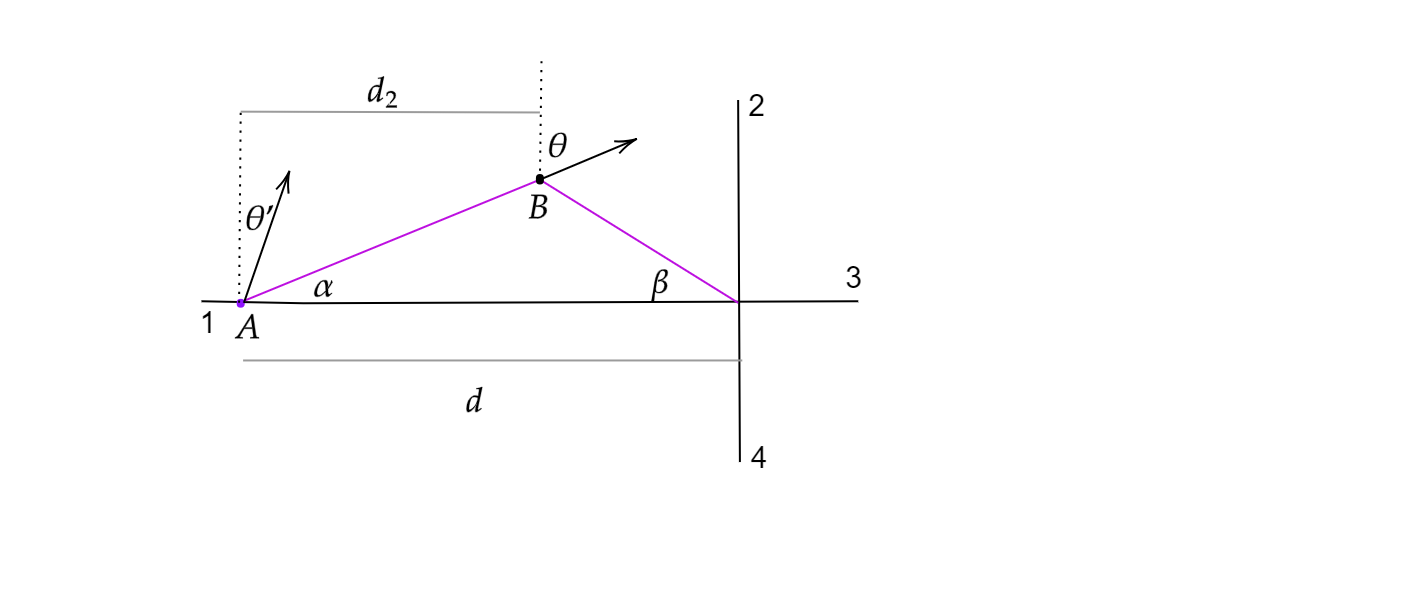}
  \caption{diagram of input in SMP 1 system}
  \label{fig:SMP 1}
\end{figure}

\begin{enumerate}
    \item If $\alpha$ and $\beta$ is below $90^o$ then output is high.
    \item If  $\Delta d$ is positive and $\alpha$ is above $90^o$ then output is low.
    \item If  $\Delta d$ is positive and $\beta$ is above $90^o$ then output is low.
    \item  If heading increment is small and  $\alpha$ and $\beta$ is below $90^o$ then output is high. 
    \item  If heading increment is small and $\Delta d$ is positive  and $\alpha$ above $90^o$ then output is low. 
    \item  If heading increment is small and $\Delta d$ is positive  and $\beta$ above $90^o$ then output is low. 
    \item  If heading increment is large and $\alpha$ and $\beta$ is below $90^o$ then output is low.
    \item If speed is zero then output is high.
    \item If $\Delta d$ is negative then output is average.
    \item If $\Delta d$ is positive then output is low.
    \item If speed is high and heading increase is small then output is average. 
    \item If speed is high and heading increase is 180 then output is high.
\end{enumerate}

FIS score above 60 indicates that the current position fix is still travelling in the previous selected link~\cite{quddus2006high}. 

\subsubsection{Subsequent Map Matching Process at a Junction (SMP2)}
\begin{figure}[h]
\centering
  \centering
  \includegraphics[width=.7\linewidth]{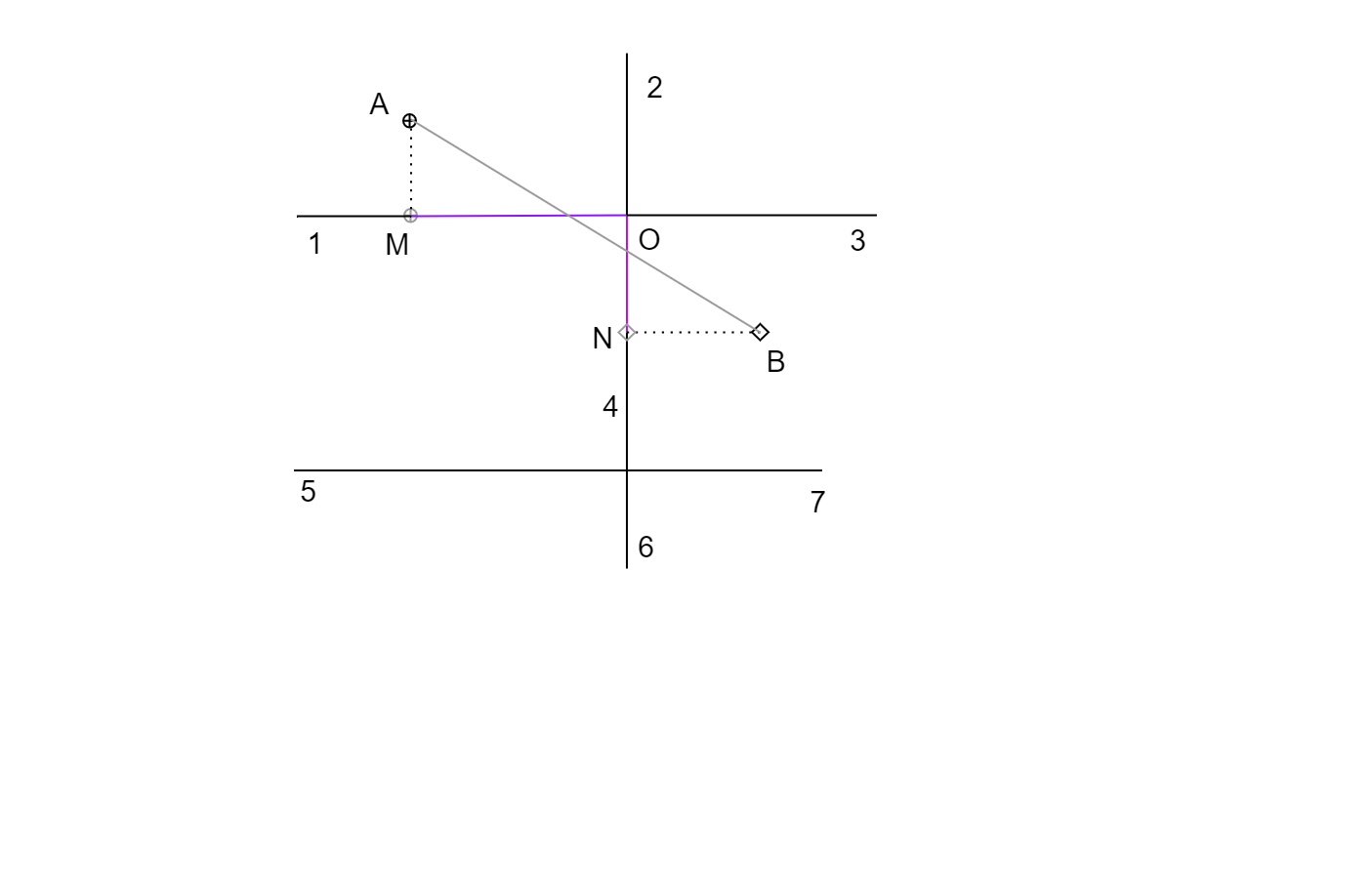}
  \vspace{-1in}
  \caption{AB is the distance travelled two position fix and MO + ON is the shortest path travelled for candidate link 4}. 
  \label{fig:SMP_2}
\end{figure}

When the FIS score obtained from SMP 1 step is less than 60, this indicates that the vehicle is entering a junction and SMP 2 is used to determine which link should be selected for the current position fix. SMP 2 algorithm is similar to IMP algorithm , but with two additional input and four additional rules \cite{gorte2014}. The two additional input are : 1) the link connectivity (1 when previous link are connected to candidate link, 0 otherwise) 2) distance error, which is defined as the difference between distance travelled by the vehicle and the shortest path travelled through road network(see \ref{fig:SMP_2}). 

The 4 additional rules that incorporated the connectivity and distance error are: 
\begin{itemize}
    \item If connectivity is low then output is low. 
    \item If connectivity is high then output is high. 
    \item If distance error is low then output is low.
    \item If distance error is high then output is high. 
\end{itemize}
Similar to the IMP process candidate link with the highest FIS score is selected as the new matched link.

\section{Implementation}

\begin{figure}[H]
    \centering
    \def\svgwidth{\linewidth}
    {\footnotesize
    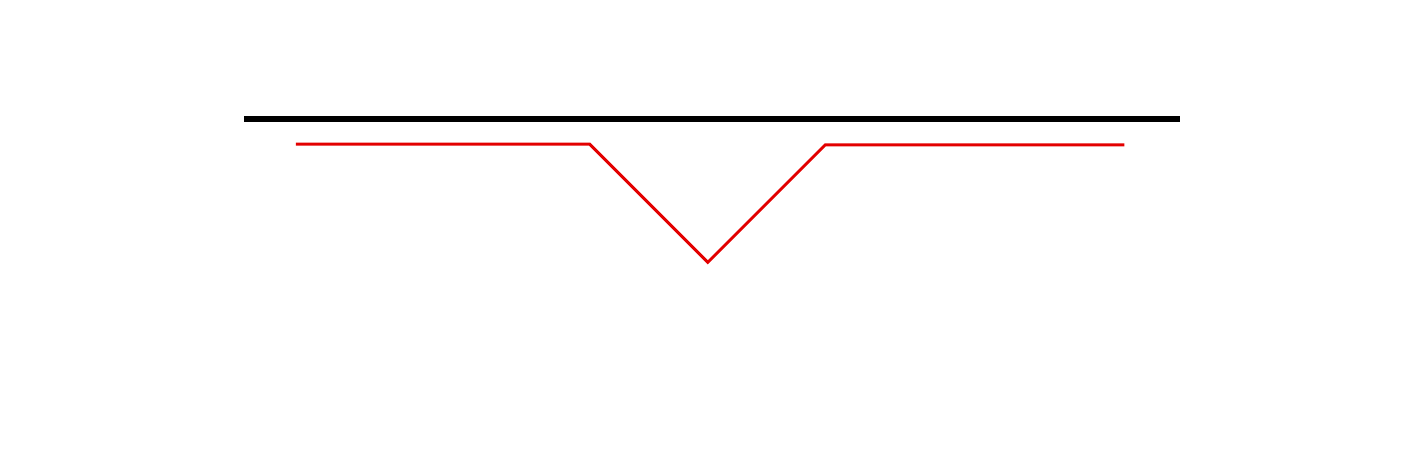}
    \footnotesize
    \begin{align*} 
        d_0 &= \text{length of ground truth} \\
        d_- &= \text{length of prediction route erroneously subtracted} \\
        d_{+} &= \text{length of prediction route erroneously added}
    \end{align*}
    \normalsize
    \caption{Error Formula by Newson and Krumm}
    \label{fig:error-formula}
\end{figure}

We implemented AHP and Fuzzy Map matching algorithms using geopandas and Osmnx libraries in Python to evaluate their performance numerically.
For both the AHP and Fuzzy, an error polygon, rather than an ellipsoid, is constructed in order to simplify our calculations. KCMMN dataset~\cite{KCMMNdataset} is used to test the performance of our algorithm. We use the following method proposed by Newson and Krumm ~\cite{newsonHiddenMarkovMap2009} : 

\begin{align*}
    \mathrm{Err} = \frac{d_- + d_{+}}{d_0},
\end{align*}
where $d_0$ is the length of the correct route, $d_-$ is the length of the prediction erroneously subtracted from the correct route, and $d_{+}$ is the length of the prediction erroneously added outside the correct route.
See figure \ref{fig:error-formula}. 

\vspace{-2.5mm}

A lower value indicates that our algorithm perform really well in matching trajectory points to a map, while a higher value indicates that our algorithms perform poorly. 
Both AHP and Fuzzy logic map matching algorithm is tested in 20 trajectories. We then compare our methods to Fast Map Matching (FMM) framework \cite{fmm}, which is based on hidden Markov model. 


\section{Results}

\subsection{Performance Evaluation}

\begin{figure}[H]
    \centering
    \includegraphics[scale = 0.4]{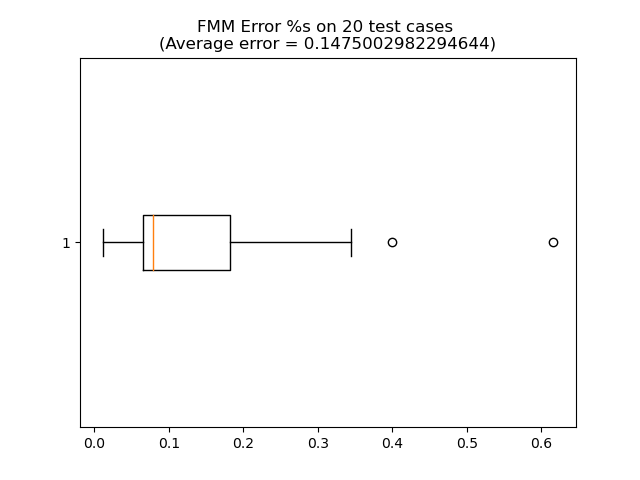}
    \caption{Performance of FMM}
    \label{fig:fmm_performance}
\end{figure}

\begin{figure}[H]
    \begin{minipage}[b]{0.48\linewidth}
        \centering
        \includegraphics[scale=0.5]{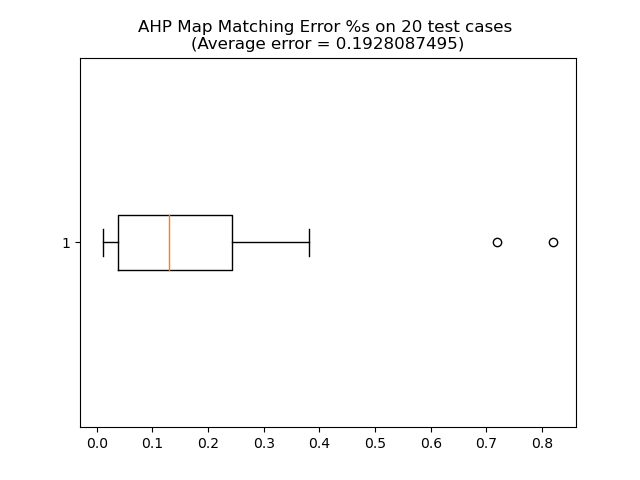}
        \caption{Performance of AHP map matching}
        \label{fig:ahpperformance}
    \end{minipage}
    \begin{minipage}[b]{0.48\linewidth}
        \centering
        \includegraphics[scale=0.5]{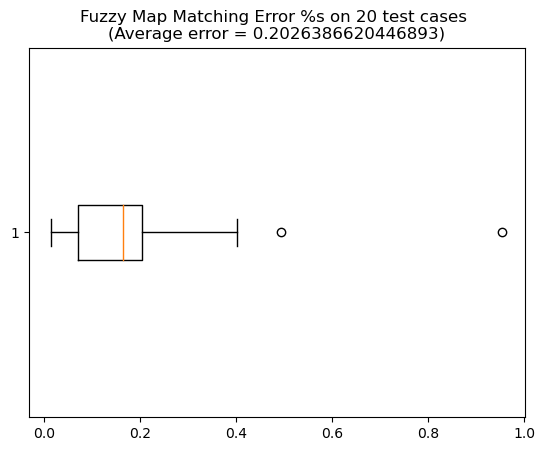}
        \vspace{0.7em}
        \caption{Performance of fuzzy logic map matching}
        \label{fig:fuzzyperformance}
    \end{minipage}
\end{figure}

From the result shown in Figure \ref{fig:fmm_performance}, \ref{fig:ahpperformance}, and \ref{fig:fuzzyperformance}, we observe that our methods perform relatively well with 19 and 20 percent error for AHP and Fuzzy logic, respectively. However both of our methods still were not able to beat FMM, which we believe due to two following reasons: data availability and parameter tuning. We used KCMMN data set to test the performance of our method because this data set included ground truth, however, KCMMN data set only has the GPS position and time, while both our methodologies require other inputs such as speed and direction data. To circumvent this problem, we estimated both speed and direction using the location data. We believe that this estimation causes our method to perform poorly on some trajectories that have more noisy measurements. Due to time limitations, we were not able to optimize the parameter that are needed for our map matching algorithm. We believe that our algorithm will perform significantly better if we are able to address these problems. 

\subsection{AHP Map Matching Results}

\begin{figure}[H]
    \centering
    \includegraphics[scale=0.7]{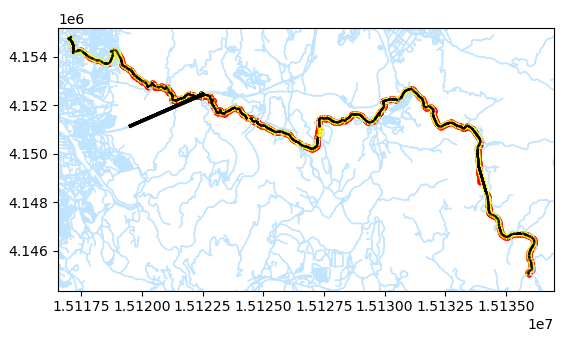}
    \caption{The result of AHP map matching for a trajectory in KCMMN dataset \cite{KCMMNdataset}, which is licensed under \href{https://creativecommons.org/licenses/by-sa/4.0/legalcode}{(CC BY-SA 4.0)}, and this figure is also licensed under \href{https://creativecommons.org/licenses/by-sa/4.0/legalcode}{(CC BY-SA 4.0)}}
    \label{fig:ahpmmresult}
\end{figure}

\begin{figure}[H]
    \centering
    \begin{minipage}[b]{0.42\linewidth}
        \includegraphics[scale=0.5]{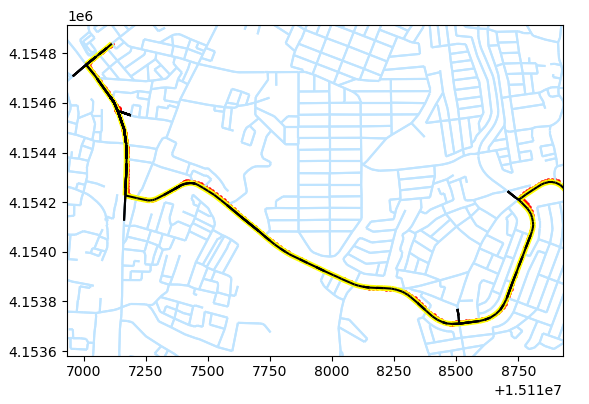}
        \caption{Some branches. (The trajectory was taken from KCMMN dataset \cite{KCMMNdataset}, which is licensed under \href{https://creativecommons.org/licenses/by-sa/4.0/legalcode}{(CC BY-SA 4.0)} and this figure is also licensed under \href{https://creativecommons.org/licenses/by-sa/4.0/legalcode}{(CC BY-SA 4.0)}.)}
        \label{fig:ahpbranch}
    \end{minipage}
    \hspace{1cm}
    \begin{minipage}[b]{0.42\linewidth}
        \includegraphics[scale=0.5]{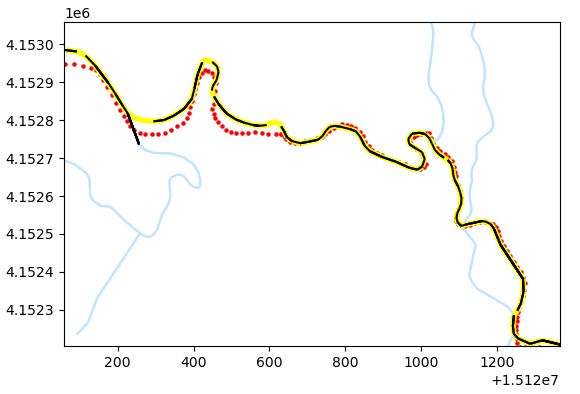}
        \caption{Some jumps. (The trajectory was taken from KCMMN dataset \cite{KCMMNdataset}, which is licensed under \href{https://creativecommons.org/licenses/by-sa/4.0/legalcode}{(CC BY-SA 4.0)} and this figure is also licensed under \href{https://creativecommons.org/licenses/by-sa/4.0/legalcode}{(CC BY-SA 4.0)}.)}
        \label{fig:ahpjump}
    \end{minipage}
\end{figure}

One of the results of AHP map matching algorithm is shown in Figure~\ref{fig:ahpmmresult}. The yellow line represents the true traveling route, the black line represents the matching result, and the red points behind the two lines represent trajectory points. Except for the big mismatch in the middle left of the figure, the entire outline looks satisfactory. However if we zoom in on local areas we can find some mismatching such as branches (Figure~\ref{fig:ahpbranch}) and jumps (Figure~\ref{fig:ahpjump}). These mismatching are more likely to occur near junctions and around points where there are many curves.

The algorithm has both advantages and disadvantages. First, it is simple and easy to understand, which helps us with the implementation. In terms of speed, it is comparable to other existing algorithms and we believe that the execution speed could be much higher through the use of other programming languages and parallelization techniques. Furthermore, the accuracy is not so bad, at least it provides results that allow us to grasp the overall outline.

As for the disadvantages, it is very sensitive to measurement errors since we use only two or three factors in each step. If any of these data is inaccurate, it could directly affect the outcomes and result in bad mismatching. Also there is a large variation in accuracy depending on the trajectory data. When the trajectory points are placed linearly, the algorithm tends to work well but when the trajectory points are curved, mismatching is more likely to occur.

To improve the algorithm, we need to find a way to deal with those mismatching. However we believe it would be challenging to avoid mismatching solely with our algorithm. Therefore, implementing post-processing will be necessary, which will be explained in Appendix~\ref{subsection:postprocessing}. 

\subsection{Fuzzy Logic Map Matching Results}
\label{section:FLMM_res}

\begin{figure}[H]
    \centering 
     \includegraphics[width=0.6\linewidth]{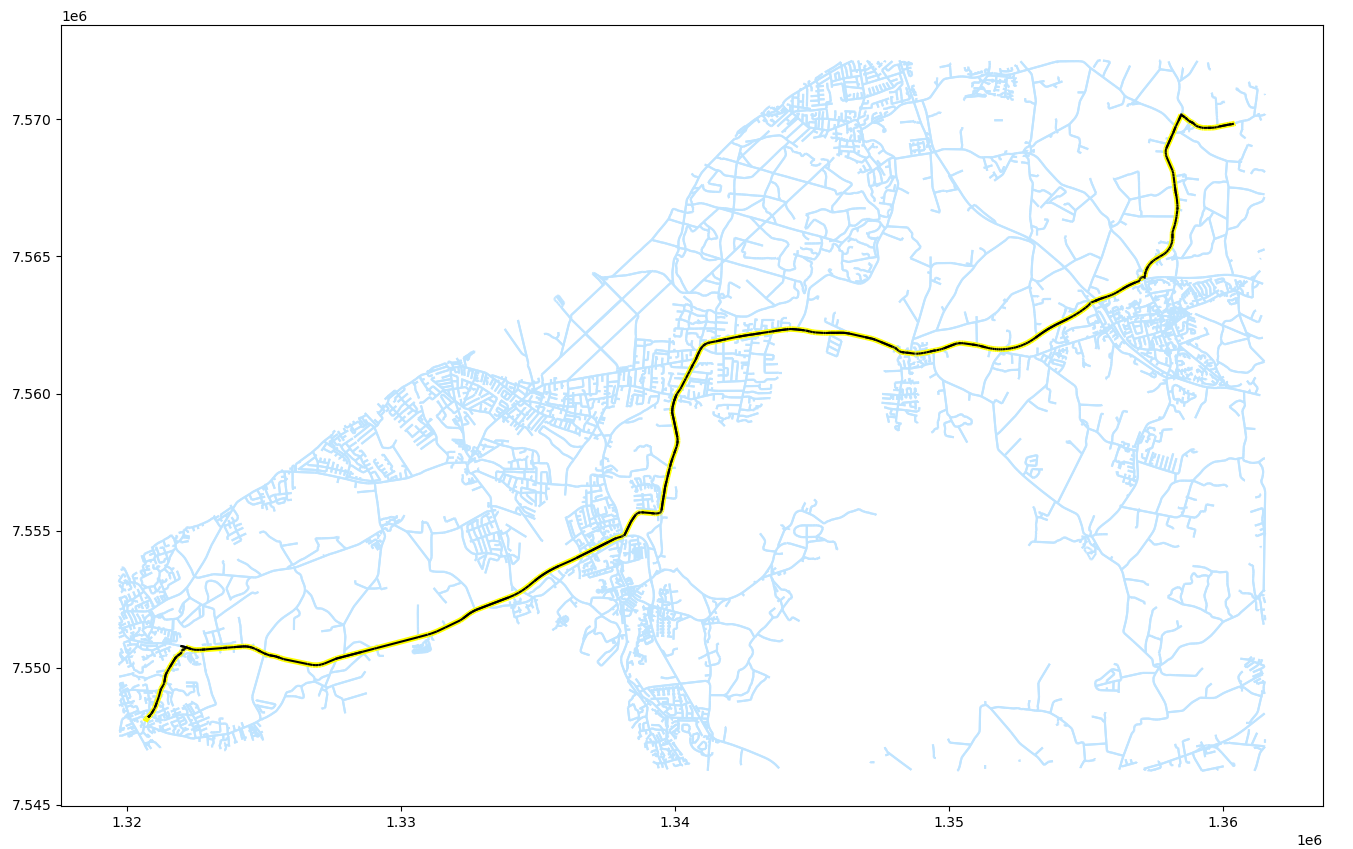}
    \caption{Fuzzy logic map matching process in KCMMN data set \cite{KCMMNdataset} licensed under \href{https://creativecommons.org/licenses/by-sa/4.0/legalcode}{(CC BY-SA 4.0)} and this figure is also licensed under \href{https://creativecommons.org/licenses/by-sa/4.0/legalcode}{(CC BY-SA 4.0)}.}
    \label{fig:Fuzzy_mm}
\end{figure}

Figure \ref{fig:Fuzzy_mm} shows an example of using fuzzy logic map matching. The yellow line represents the ground truth and the black line represents the road selected by our algorithm. Similar to AHP, Fuzzy logic Map Matching also suffers from branches and jumps that occur due to selecting incorrect links. 

\begin{figure}[H]
    \centering
    \begin{minipage}[c]{0.45\textwidth}
        \includegraphics[width=10cm,trim={6cm 0 0 0},clip]{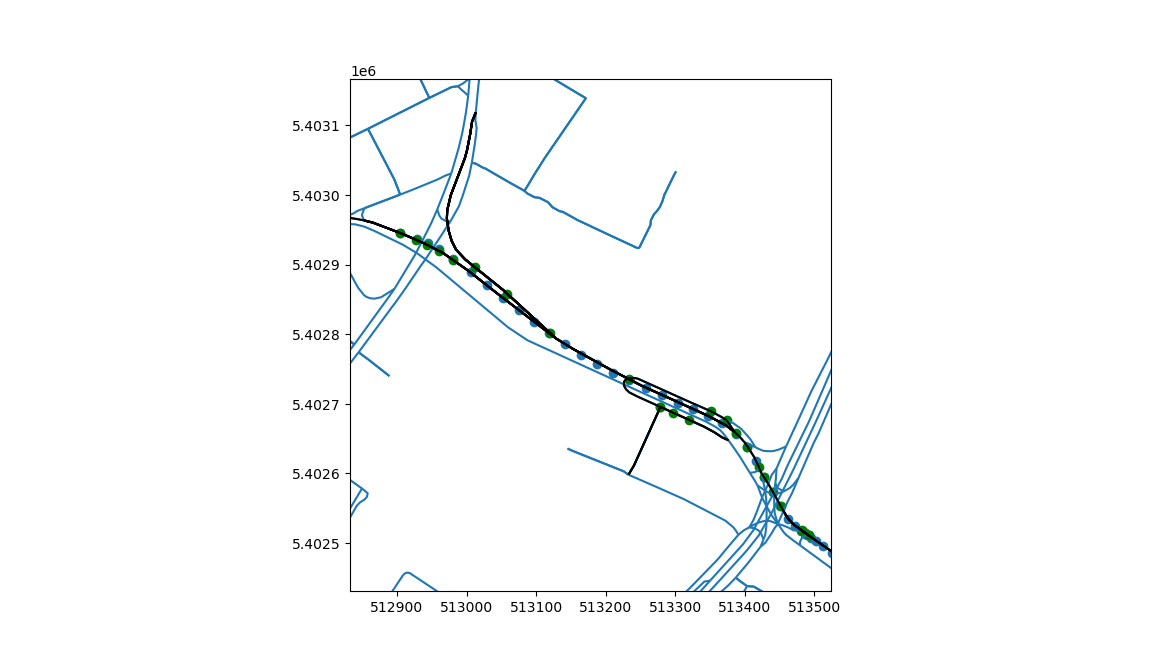}
        \vspace{-10pt}
        \captionof{figure}{Before. (The trajectory was taken from KCMMN dataset \cite{KCMMNdataset}, which is licensed under \href{https://creativecommons.org/licenses/by-sa/4.0/legalcode}{(CC BY-SA 4.0)} and this figure is also licensed under \href{https://creativecommons.org/licenses/by-sa/4.0/legalcode}{(CC BY-SA 4.0)}.)}
        \label{fig:flmm_before}
    \end{minipage}
    \hspace{1cm}
    \begin{minipage}[c]{0.45\textwidth}
        \includegraphics[width=10cm,trim={10cm 0 0 0},clip]{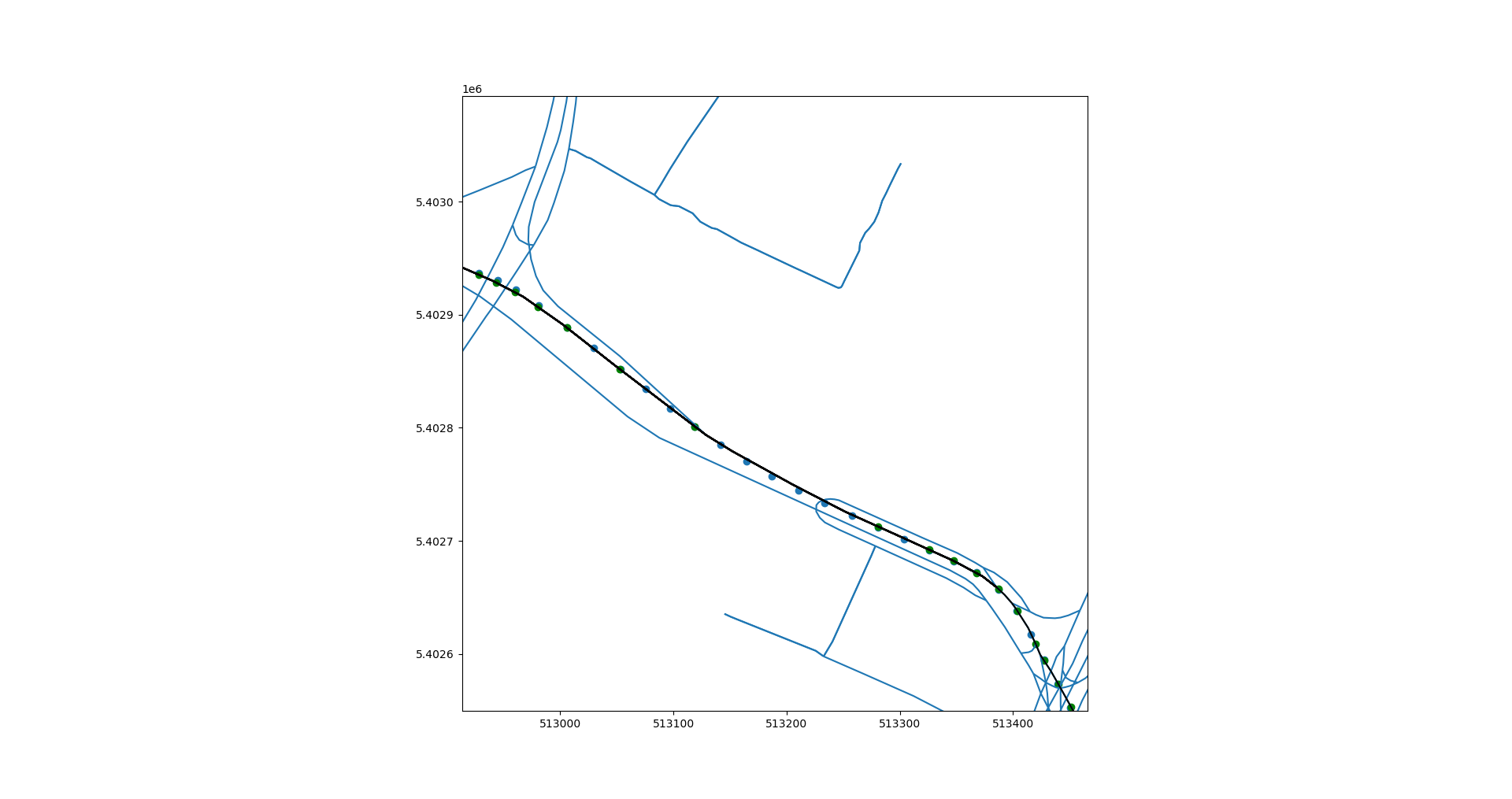}
        \vspace{-10pt}
        \captionof{figure}{After. (The trajectory was taken from KCMMN dataset \cite{KCMMNdataset}, which is licensed under \href{https://creativecommons.org/licenses/by-sa/4.0/legalcode}{(CC BY-SA 4.0)} and this figure is also licensed under \href{https://creativecommons.org/licenses/by-sa/4.0/legalcode}{(CC BY-SA 4.0)}.)}
        \label{fig:flmm_after}
    \end{minipage}
\end{figure}

We introduce a rule weight $a_i$ to indicate how confident we are with certain rules. For example in our KCMMN implementation, since we know that speed and velocity are estimated rather than measured, we can put higher weight on connectivity and perpendicular distance and less on speed and velocity. Figure \ref{fig:flmm_before} shows the results of Fuzzy logic map matching before we implemented the rule weight. The blue line represents the GPS trajectory data, the black line represents which link is selected and the green point represents where the trajectory point is matched to the selected link. We observe that some of the trajectory points are matched to the wrong link that caused the branches to occur. Figure \ref{fig:flmm_after} shows the results after we implemented the rule weight in our FIS system, where we observe that the trajectory is matched correctly even in the complicated junctions. 

During our implementation of fuzzy logic map matching, we noticed that although this method performs as well as other methods, it is less sensitive to error when some subsets of the input are miss-specified. The fuzzy logic map matching error rate increases by 10 percentage point when the direction was incorrectly specified, which is relatively lower than our other method that differs by 60 percentage point. 

One disadvantage of Fuzzy Logic map matching is that the computational time is relatively slower than both FMM and AHP methods. One possible solution is to implement this algorithm in C manually in order to accelerate the computing time. 
\vspace{-10pt}

\section{Appendix}

\subsection{Preprocessing by DBSCAN}\label{subsection:dbscan}
In this subsection, we talk about stay point mitigation and outlier detection by DBSCAN, and a method to automatically determine a parameter of the algorithm.

DBSCAN is a clustering algorithm based on metric information.
The algorithm has two parameters: \variable{minPts} and \variable{eps}.
Then the algorithm takes a set of points in space as its input and classifies them into three categories: core points, reachable points, and unreachable points.
Among these classes, core points and reachable points belong to a cluster, meanwhile, unreachable points do not belong to any cluster.
The \variable{minPts} parameter is the density threshold for a point to become a core point, and the \variable{eps} parameter is the search radius for each point.
In ~\cite{jafarlou_improving_nodate}, they used DBSCAN to detect and mitigate stay points by replacing each cluster with a single point.
Here, inspired by their method, we propose to utilize DBSCAN to detect outliers simply by labeling the unreachable points of DBSCAN as outliers.
Since many map matching algorithms depend on the number of trajectory points for their computation time, conducting DBSCAN as preprocessing would speed up those algorithms.

Although these methods of stay point mitigation and outlier detection are simple, the choice of the parameters of the algorithm is critical when actually conducting the algorithm as preprocessing of the map matching algorithm.
If the \variable{eps} parameter is too small, too many points may be classified as unreachable points.
On the other hand, if the \variable{eps} parameter is too large, an undesirably large portion of points may be grouped into a single cluster.
As for the \variable{minPts} parameter, in~\cite{sander1998density} they suggested setting it to $2 \times d$, where $d$ is the dimension of the space.
As for the \variable{eps} parameter, though such a simple way to determine the value is not known unlike \variable{minPts}, the following heuristic based on "elbows" is said to be a good way~\cite{ester1996density}:
\begin{enumerate}
    \item Prepare an empty list \variable{l};
    \item For each point in the input set, compute the distance between the point and the \variable{minPts}-th closest point from it, and put it into \variable{l};
    \item Sort \variable{l} in ascending order;
    \item Plot the set $\{(i, \mathbf{l[i]}) : 0 \leq i < \variable{len}(\variable{l})\}$ in a plane;
    \item Consider a curve that interpolates the plotted points;
    \item The elbows, where the curve rapidly goes up, are considered to be good candidates for the \variable{eps} parameter.
\end{enumerate}
When one tries this heuristic, it poses a problem how to determine the interpolation curve and elbows.

To deal with this problem, we propose an elbow detection algorithm in Algorithm ~\ref{elbow_detection}.
\begin{algorithm}

\KwIn{List \variable{l} of the \variable{minPts}-distances, degree $\theta$, integer $d$}
\KwOut{Set of real numbers}
Set $\mathbf{points} = \{\{(i, \mathbf{l[i]}) : 0 \leq i < \variable{len}(\mathbf{l})\}\}$;
Rotate \variable{points} $-\theta$ degree around the origin;

Compute the least square polynomial fit of degree $d$ of the rotated points;

Find the local minimals $m$ of the fitting polynomial;

Rotate $m$ $\theta$ degree around the origin;

Extract points whose $x$-coordinate is in $[0, \mathbf{len}(\mathbf{l}))$ from $m$, and return their $x$-coordinates as the output;
\caption{Elbow Detection Algorithm}
\label{elbow_detection}
\end{algorithm}

Although this algorithm works well for GPS trajectory points as far as we manually checked with our eyes, it is desired that it is verified in various different tasks.
Also, because we tested only the least square polynomial fit as the smooth curve which interpolates the points in this research, it may be possible that other methods are superior to it.
Therefore, it remains as tasks that should be examined in the future to develop a method to find proper values for $\theta$ and $d$, and to test other methods of the interpolation curve.

\subsection{Postprocessing}\label{subsection:postprocessing}
As we have seen in the sections of the AHP and fuzzy-logic map matching algorithms, to achieve greater accuracy, it is necessary to deal with many "jumps" and "branches" that the route predicted by these algorithms includes.
We suggest after the map matching algorithm applying postprocessing to the predicted route so as to interpolate the jumps and remove the branches.
Specifically, we propose the following procedure:
\begin{enumerate}
    \item Find the shortest path from the start point to the end point of the predicted route;
    \item Subtract the edges in the shortest path from the set of the edges in the predicted route by the map matching algorithm;
    \item Remove all connected components of the shape of a linear graph.
\end{enumerate}

The reason why the removed connected components have to be linear graphs in step 3 is that the correct route may have some loops in it, and thus, we do not want to mistakenly remove them.

\printbibliography[title={References},heading=bibintoc]

\end{document}